\documentclass[galaxies,article,accept,pdftex,moreauthors]{Definitions/mdpi}

\firstpage{1}
\pubvolume{1}
\issuenum{1}
\articlenumber{0}
\pubyear{2024}
\copyrightyear{2024}
\datereceived{30 April 2024}
\daterevised{23 May 2024} 
\dateaccepted{25 May 2024}
\datepublished{ }
\hreflink{\linebreak} 

\usepackage{bm} 
\usepackage{makecell}

\Title{The Response of the Inner Dark Matter Halo to Stellar Bars}

\TitleCitation{The response of the inner dark matter halo to stellar bars}

\Author{Daniel A. Marostica$^{1}$\orcidA{}, Rubens E. G. Machado$^{1}$\orcidB{}*, E. Athanassoula$^{2}$\orcidC{} and T. Manos$^{3}$\orcidD{}}

\AuthorNames{Daniel A. Marostica, Rubens E. G. Machado, E. Athanassoula and T. Manos}

\AuthorCitation{Marostica, D. A.; Machado, R. E. G; Athanassoula, E.; Manos, T.}

\address{%
$^{1}$ \quad Departamento Acad\^emico de F\'isica, Universidade Tecnol\'ogica Federal do Paran\'a, Av. Sete de Setembro 3165, Curitiba, Brazil\\
$^{2}$ \quad Aix Marseille Univ, CNRS, CNES, LAM, Marseille, France\\
$^{3}$ \quad ETIS Lab, ENSEA, CNRS, UMR8051, CY Cergy-Paris University, Cergy, France\\
} 

\corres{Correspondence: rubensmachado@utfpr.edu.br}

\abstract{Barred galaxies constitute about two thirds of observed disc galaxies. Bars affect not only the mass distribution of gas and stars, but also that of the dark matter. An elongation of the inner dark matter halo is known as the halo bar. We aim to characterise the structure of the halo bars, with the goal of correlating them with the properties of the stellar bars. We use a suite of simulated galaxies with various bar strengths, including gas and star formation. We quantify strengths, shapes, and densities of these simulated stellar bars. We carry out numerical experiments with frozen and analytic potentials in order to understand the role played by a live responsive stellar bar. We find that the halo bar generally follows the trends of the disc bar. The strengths of the halo and stellar bars are tightly correlated. Stronger bars induce a slight increase of dark matter density within the inner halo. Numerical experiments show that a non-responsive frozen stellar bar would be capable of inducing a dark matter bar, but it would be weaker than the live case by a factor of roughly two.}

\keyword{galactic dynamics; barred galaxies; numerical simulations}

\begin{document}

\section{Introduction}

Observations indicate that bars are present in most galaxies in the local Universe, with an often quoted fraction of about two thirds \citep[e.g.][]{2000AJ....119..536E, Sheth_2008, 2010ApJ...714L.260N, 2011RMxAC..40..120G, 2012ApJ...757...60K, 2019ApJ...872...97L}. Bars are persistent long-term asymmetries which represent a rich source of information about galaxy structure and dynamics. Many studies, from isolated objects to large surveys, have added valuable information about their properties, including morphological \citep[e.g.][]{1979ApJ...227..714K, 2009AJ....137.4487B, 2012MNRAS.423.3486W, Ciambur2017, Mendez2018}, kinematical \citep[e.g.][]{Nitschai2019} and photometrical ones \citep[e.g.][]{2011MNRAS.415.3308G, 2020A&A...641A.111C}. Some of the observed effects of bars include changes in star formation rates \citep[e.g.][]{2011MNRAS.416.2182E, 2012MNRAS.424.2180M, 2013ApJ...779..162C}, metallicity \citep[e.g.][]{Williams_2012}, age gradients \citep[e.g.][]{2019MNRAS.484.5192C} and active galaxies \citep[e.g.][]{2004ApJ...607..103L}.

Simulations have long been employed to model the structure and dynamics of barred galaxies. Since the end of the last century, papers dealt with fundamental features of bars and their relations with the environment, as well as with dark matter \citep[e.g.][]{1996ASPC...91..309A, 1996ASPC...91..357D, 1998ApJ...493L...5D, 1998MNRAS.300...49B, 2000ApJ...543..704D, AthanassoulaMisiriotis2002, Athanassoula2003, 2009ApJ...703.2068D, 2010ApJ...719.1470V}.

In the past decade, several dynamical issues on the evolution of bars have been explored by means of simulations of isolated galaxies. For instance, \citet{2015MNRAS.450..229F} assessed the effect of boxy/peanut bulges in hydrodynamical simulations, showing that these structures are anti-correlated with bar strength. \citet{Debattista2017} showed that, when bars are formed, populations of stars are gradually separated, potentially building a bulge. \citet{Marasco2018} confirmed previous findings \citep{Athanassoula2003} that bar-like patterns are common even in galaxies where baryons are not dominant. \citet*{Athanassoula2013} -- hereafter AMR13 --  focused on the effects of gas on the formation of stellar bars. The formation of bars under interactions with other galaxies due to tidal effects has also been explored via simulations \citep[e.g.][]{2014MNRAS.445.1339L, 2014ApJ...790L..33L, Lokas2016, Lokas2018, Gajda2018}.

The formation of realistic galaxies in high-resolution fully cosmological simulations has also become possible in recent years. For example, \citet{Scannapieco2009} described the formation of discs in haloes from the Aquarius Simulation and \citet{2012MNRAS.425L..10S} studied the first spontaneously-formed bars in a $\Lambda$CDM universe. More recently, \citet{Algorry_2017} looked into the formation and evolution of bars in the EAGLE Simulation \citep{2015MNRAS.446..521S}, finding that strong bars form quickly in disc-dominated, gas-poor systems with declining rotation curves. \citet{Guevara2020} performs a similar analysis with barred galaxies from the IllustrisTNG simulation \citep{2014MNRAS.445..175G}, reporting quantitative bar features in good agreement with observations and with simulations of isolated galaxies.

It is inferred from such cosmological simulations that dark matter halos are usually prolate, as many works have indicated \citep[e.g.][]{Frenk1988, Dubinski1991, Warren1992, Cole1996, Jing2002, Bailin2005, Allgood2006, Novak2006, Bett2007}. \citet{Velliscig2015} and \citet{Chua2019} have obtained the same results by analysing EAGLE and Illustris simulations, respectively. Both cosmological simulations and numerical experiments have generally indicated that a baryonic disc has the effect of partially washing out the prolateness or triaxiality of the dark matter halo \citep[e.g.][]{Berentzen2006, Abadi2010, Tissera2010, Machado2010, Artale2019, Cataldi2021}. Recently, \citet{2019MNRAS.tmp.2477P} compared halo shapes of a set of galaxies from the Auriga simulation \citep{2017MNRAS.467..179G} and also found that the haloes from the hydrodynamical simulation were rounder than those from the dark-matter-only simulation. Thus the general picture is that the baryons tend to circularise the halo.

Quantifying how dark matter responds to the presence of the stellar disc is important to understand their joint dynamical evolution. From the theoretical point of view, \citet{1984MNRAS.209..729T, 1985MNRAS.213..451W} used Hamiltonian formalism to explore the perturbations due to a rotating bar potential.

A strongly barred disc may again drive the inner dark matter halo into an elongated shape. The so-called `halo bar', the main object of this paper, was first discussed by \citet{1992ApJ...400...80H}, who detected this structure using a live halo and a rigid disc bar. Their work was then followed by others which used higher resolution collisionless $N$-body simulations. 

As examples, \citet{ONeill2003} briefly describe an elongation in the inner halo in synchronous rotation with the disc bar, partly responsible for slowing it down. \citet{Athanassoula2005} uses live discs and haloes to study their structure and dynamical interplay, finding that the halo bar is prolate-like and increasingly axisymmetric as one moves away from its center. \citet{Colin2006} analyse the synchrony of the pattern speed and the growth rate of the bars, finding that a strong correlation exists between them. \citet{Berentzen2006} call the halo bar a `ghost bar', arguing that this gravitational wake is so dependent on the structure and dynamics of the disc bar that, in the absence of the latter, it would eventually disappear. \citet{Athanassoula2007} does one of the first thorough analyses of the halo bar using fiducial models of high-resolution simulations, finding that the halo bar is always much shorter than the stellar bar and that both turn with roughly the same pattern speed. \citet{Petersen2016} find that the stellar disc is responsible for trapping dark matter particles by transferring angular momentum, thus inducing the formation of a `shadow bar', with forced orbits within the region of the halo bar. This kind of structure has also been noted in other works \citep{ONeill2003, HolleyBockelmann2005, Debattista2008} and some authors have studied the implications of the halo bar for dark matter detection experiments \citep{2005PhRvD..72h3503A, 2010JCAP...02..012L, Petersen2016D, Kavanagh2016}. Studying regular and chaotic orbits, \citet{Machado_2016} found that the halo bar was the only region where chaotic orbits were more numerous than regular ones in the early evolution.

AMR13 explored the formation and evolution of barred galaxies in the presence of both gas and/or triaxial dark matter haloes. In that paper, a suite of simulations was used to analyse the morphology, bar strength and numerous other properties of both the disc and the halo. Some of the main results were that gas-rich discs inhibit the formation of strong bars. Similarly, strongly triaxial haloes tend to lead to weaker bars. Regarding the halo bar itself, AMR13 found that it also develops in simulations including a gaseous disc and that it must be linked to the presence of a strong bar in the stellar component. These simulations were also used to study the imprints of boxy/peanut bulges on the 2D line-of sight kinematics, in order to help identify the existence and properties of such structures in Integral Field Unit observations of disc galaxies \citep{2015MNRAS.450.2514I}.

The hydrodynamical $N$-body simulations in AMR13 take into account not only the presence of a gaseous component in the discs, but they also include star formation, feedback and cooling. As a result, these simulations constitute a diverse sample of realistic galaxies with varying bar strengths. This data set is well suited for the systematic exploration which is the goal of the present paper. Here, we aim to extend the study of the halo bars by systematically quantifying their structure, with the goal of understanding how the inner dark matter halo responds to the stellar bar. To this end, we analyse the shape, strength and density of the halo bar. We present additional numerical experiments with unresponsive discs to investigate the role of the stellar bar.

This paper is organized as follows. In section~2 we describe the set of simulated galaxies. In section~3 we present the results of our analyses. We conclude with a summary and discussion in section~4.

\section{Simulations}

In this paper, we will measure the properties of galaxies from the suite of hydrodynamical $N$-body simulations presented in AMR13. Here we briefly summarise only the main features of those simulations which are needed to present our results. For further details about the simulation setup, the reader is referred to that paper and references therein.

The simulations in AMR13 were carried out for 10\,Gyr using a version of the \textsc{gadget2} \citep{Springel2005} including star formation. There are 15 galaxies, with all the combinations of 3 initial halo shapes and 5 initial gas fractions. The initial halo shapes are: spherical ($b/a=1$, $c/a=1$), triaxial ($b/a=0.8$, $c/a=0.6$), and more triaxial ($b/a=0.6$, $c/a=0.4$). The halo mass is $M_{\rm h}=2.5\times10^{11}\,{\rm M_{\odot}}$ and it is always the same in all 15 models. The disc mass is $M_{\rm d}=5\times10^{10}\,{\rm M_{\odot}}$ and it is always the same in all 15 models. The disc component is initially composed of so-called old stars and gas. The initial gas fractions range from 0\% to 100\% in the initial conditions. As the simulation evolves, gas is converted into new stars. By about $\sim$2\,Gyr, the gas fraction has dropped to below 20\% in all models. This has profound impacts on the formation of the bar, which are discussed in AMR13.

\begin{figure}
    \includegraphics[]{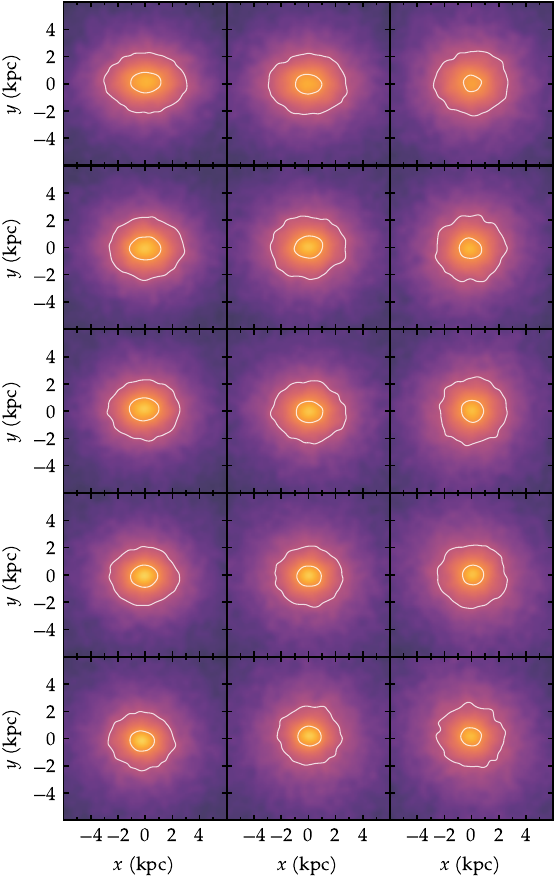}
    \caption{Projected dark matter mass of halo particles within $|z|<1$\,kpc at $t=10$\,Gyr. The contours highlight the shape of the inner halo. As in Table~\ref{tab1}: columns are initial halo shapes; rows are initial gas fraction.}
    \label{xy101}
\end{figure}

Table~\ref{tab1} gives the labels of the 15 models and the initial conditions. The 3 columns display the halo triaxialities; the 5 rows display the initial gas fractions. This is also the $3\times5$ layout of other figures in this paper. For instance, in Figure~\ref{xy101} we present the projected dark matter mass on the $xy$ plane of halo particles within $|z|<1$\,kpc. For the Fourier analyses in this paper, the halo will be limited to this equatorial slice, because projecting halo particles from all heights would attenuate the measurements \citep{Athanassoula2007}. The snapshots in Figure~\ref{xy101} are from the end of the simulation ($t=10$\,Gyr) and the white contours help highlighting the general shape of the inner halo. Strong bars tend to be towards the top left of Figure~\ref{xy101} (model 101) and weak bars towards the bottom right (model 121). This happens because both gas content and triaxiality tend to inhibit strong bars, as discussed in AMR13. Therefore, the spherical halo with no gas hosts the strongest bar.

\begin{table}[h]
    \caption{Initial conditions and labels of the 15 models used in this work (from \citealt{Athanassoula2013}). Gas fraction is the total mass of gas particles divided by the total disc mass (gas+stars). The axis ratios $b/a$ and $c/a$ are intermediate-to-major and minor-to-major ratios.}
    \begin{tabular}{cccc}
        \hline
        \makecell{initial gas fraction} & \makecell{halo 1\\$b/a=1.0$\\$c/a=1.0$} & \makecell{halo 2\\$b/a=0.8$\\$c/a=0.6$}  & \makecell{halo 3\\$b/a=0.6$\\$c/a=0.4$}\\
        \hline
        0.00 & 101 & 102 & 103 \\
        0.20 & 106 & 109 & 110 \\
        0.50 & 111 & 114 & 115 \\
        0.75 & 116 & 117 & 118 \\
        1.00 & 119 & 120 & 121 \\
        \hline
    \end{tabular}
    \label{tab1}
    \par
    \end{table}

The techniques used to measure each property will be explained or referenced at the beginning of the subsections of section~3. The number of particles in the simulations are given in Table~\ref{tab2}.

\begin{table}
\caption{Number of particles in the simulations, for each component.}
\begin{tabular}{cccccc}
\hline
model & $N_{\rm halo}$ & $N_{\rm disk}$ & $N_{\rm gas}$ & $N_{\rm gas}$ & $N_{\rm stars}$ \\
 & & & (initial) & (final) & (final)\\
\hline
101 & 1\,000\,000 & 200\,000 &  \,   \,  0 &   \,  0 &  \,   \,  0 \\
102 & 1\,000\,000 & 200\,000 &  \,   \,  0 &   \,  0 &  \,   \,  0 \\
103 & 1\,000\,000 & 200\,000 &  \,   \,  0 &   \,  0 &  \,   \,  0 \\[0.25em]
106 & 1\,000\,000 & 160\,000 &  \,200\,000 & 40\,800 &  \,338\,857 \\
109 & 1\,000\,000 & 160\,000 &  \,200\,000 & 38\,616 &  \,341\,887 \\
110 & 1\,000\,000 & 160\,000 &  \,200\,000 & 39\,589 &  \,340\,860 \\[0.25em]
111 & 1\,000\,000 & 100\,000 &  \,500\,000 & 64\,000 &  \,906\,399 \\
114 & 1\,000\,000 & 100\,000 &  \,500\,000 & 64\,123 &  \,907\,027 \\
115 & 1\,000\,000 & 100\,000 &  \,500\,000 & 67\,816 &  \,905\,424 \\[0.25em]
116 & 1\,000\,000 &  50\,000 &  \,750\,000 & 80\,472 & 1\,383\,891 \\
117 & 1\,000\,000 &  50\,000 &  \,750\,000 & 81\,486 & 1\,384\,287 \\
118 & 1\,000\,000 &  50\,000 &  \,750\,000 & 79\,518 & 1\,387\,188 \\[0.25em]
119 & 1\,000\,000 &      \,0 & 1\,000\,000 & 94\,528 & 1\,866\,322 \\
120 & 1\,000\,000 &      \,0 & 1\,000\,000 & 94\,332 & 1\,867\,179 \\
121 & 1\,000\,000 &      \,0 & 1\,000\,000 & 94\,567 & 1\,867\,718 \\[0.25em]
\hline
\end{tabular}
\label{tab2}
\end{table}

\section{Analysis and results}
\subsection{Bar strength}

In order to quantify the strength of the bars, we adopted the usual definition (such as in AMR13). Taking the Fourier decomposition of the mass distribution:
\begin{equation}
a_m (R) = \sum _{i=0}^{N_{R}}~m_{i}~\cos (m\theta_i), ~ m=0, 1, 2, ...
\end{equation}
\begin{equation}
b_m (R) = \sum _{i=0}^{N_{R}}~m_{i}~\sin (m\theta_i), ~ m=1, 2, ... 
\end{equation}
and defining the relative amplitude of the $m$=2 component as a function of cylindrical radius $I_2 (R)$:
\begin{equation} \label{eq:Am}
I_2 (R)= \left(\frac{\sqrt{a_2^2+b_2^2}}{a_0}\right),
\end{equation}
we then define the strength of the bar as $A_2$, which is the maximum amplitude of $I_2(R)$:
\begin{equation} \label{eq:Amb}
A_2 = \max(I_2 (R)).
\end{equation}
Thus, for each instant of time, we have one $A_2$ value.

\begin{figure}[ht]
	\includegraphics[width=0.7\columnwidth]{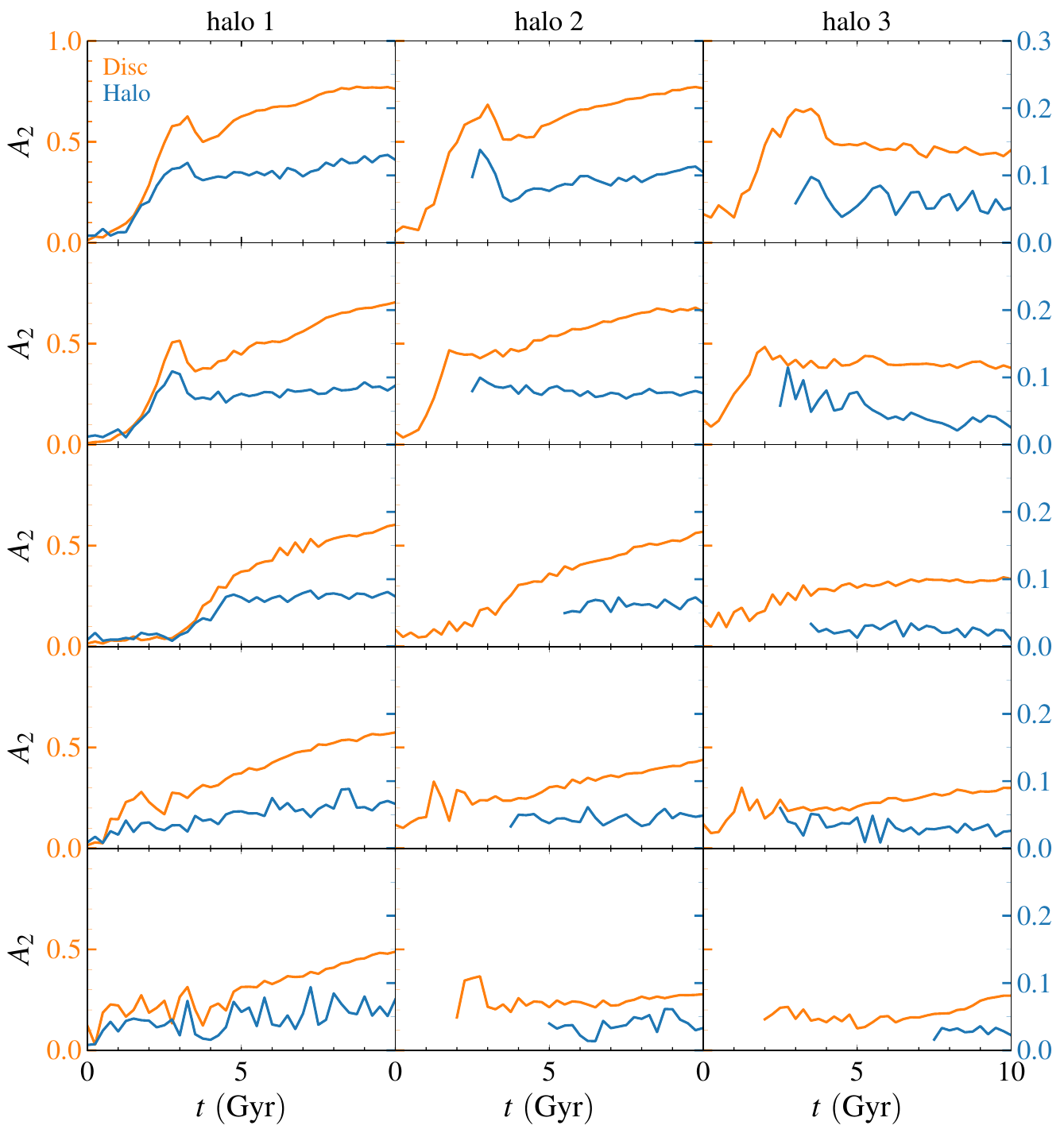}
    \caption{Strength of the disc and halo bars as a function of time. The three columns correspond to the three different initial halo triaxialities, respectively from left to right. The five rows, from top to bottom, correspond to increasing initial gas fraction (see Table \ref{tab1}). The orange curves represent the disc bar and their scale is shown in the left ticks. For better comparison, we use a different scale for the halo bar (measured within $|z| < 1$\,kpc), depicted in blue, to the right.}
    \label{strength}
\end{figure}

We measured the strengths of the disc bar and of the halo bar using the same method. In this subsection, we present the results of these measurements for all galaxies and also compare the strength of the disc bar and the halo bar at $t$=10\,Gyr. Because in the Fourier analysis the mass is projected on the plane, we have limited the vertical coordinates of the halo particles to be analysed to $|z| < 1$\,kpc. At early times, measuring $A_2$ for haloes 2 and 3 would not return information about a bar (it would be too early for the bar to have been formed), but actually about the initial triaxiality of these haloes. We have inspected each one of the $I_2(R)$ curves seeking to find meaningful peaks in the inner parts. For these reasons, parts of the $A_2(t)$ curves in Figure~\ref{strength} for the disc bars have been omitted until we could be certain that they would reflect the actual bar, rather than the initial disc ellipticity. Halo 1, however, is initially spherical and the chosen methods of measurement are able to detect small bar strengths.

Figure~\ref{strength} comprises all the 15 simulations in a five-row, three-column plot. The 5 rows correspond to initial gas fractions of 0, 20, 50, 75 and 100\%, from top to bottom. From left to right, the three columns correspond to the initial triaxialities, from halo 1 (spherical) to halo 3 (most triaxial). See Table \ref{tab1}. It allows us to compare the evolution of bar strengths through the $A_2$ curves as a function of time. The curves were smoothed with a Savitzky-Golay filter. The strength of the disc bar, plotted in orange, clearly contrasts with that of the dark matter halo, which, in order to show best the similarities among their features, has been depicted in a different scale (in blue). Notice that the vertical scale for the halo $A_2$ is stretched by a factor of about 3.

\begin{figure}[!ht]
	\includegraphics[width=0.7\columnwidth]{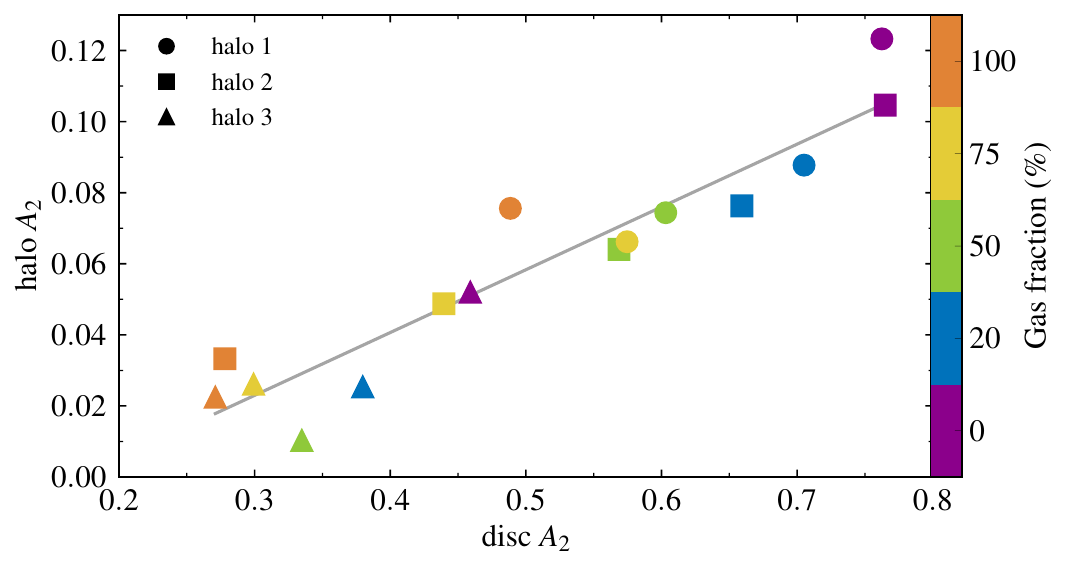}
    \caption{Halo bar strength as a function of disc bar strength at $t$ = 10\,Gyr. Circles, squares and triangles stand for haloes 1, 2 and 3, respectively. Colour is according to initial gas fraction. Scales of halo $A_2$ and disc $A_2$ differ by approximately an order of magnitude.}
    \label{A2_ratio_10gyr}
\end{figure}

We notice that the disc bar gets stronger as the simulations evolve, in some cases undergoing a period of buckling instabilities. 
Following that, the bars start to strengthen again, although, in some cases of strong triaxiality, this effect is hindered. We have thus recovered the results presented in AMR13 for the stellar bar. We now turn to the detailed analysis of the strengths of the halo bars. We find that the halo bar tends to mimic the disc bar behaviour in some cases, though in much smaller scales. However, for weak bars, this parallel evolution is less clear. For the spherical halo (first column of Figure~\ref{strength}), the early period of bar formation may be said to be very much synchronised in galaxies 101, 106 and 111. For these galaxies at least, the early growth begins not only at the same time, but also with the same rate. For the other halo shapes, it would be difficult to disentangle early halo bar growth from the intrinsic halo triaxiality. Once the halo bar has begun its growth, and particularly after the buckling, the slope of the halo $A_2$ curve is generally much flatter than that of the disc $A_2$ -- with the possible exception of the two strongest bars, 101 and 102. Towards the bottom right of Figure~\ref{strength}, the halo bars are nearly undetectable. We may set a threshold for the halo bar of approximately $A_2 > 0.05$ in order to consider it existent. In this sense, galaxies 116, 117 and 118 have extremely weak halo bars whereas those of 119, 120 and 121 barely exist. Quantitatively, we find that disc bars with $A_2 < 0.4$ at the end of their evolutions are not accompanied by relevant halo bars.

Figure~\ref{A2_ratio_10gyr} displays the final values of the halo $A_2$ curves ($t$ = 10\,Gyr) as a function of those of the disc. This represents the correlation in strength of the disc and the halo bars at the end of their evolution. The colour bar represents the initial gas fraction of each galaxy. Circles, squares and triangles represent halo 1, halo 2 and halo 3, respectively. It is important to highlight that the vertical axis scale is about ten times smaller than the horizontal one. The main feature of this plot is the approximately linear correlation in bar strength between the disc and the halo, which does not depend strongly on initial gas fraction nor halo shape at first glance. It yields, however, a Spearman's rank correlation coefficient of $\rho = 0.93$, $p = 4.24\times10^{-7}$ and a Pearson correlation coefficient of $r = 0.94$, $p = 1.7\times10^{-7}$.

\subsection{Axis ratios}
We now analyse the evolution of the halo bars by comparing their shapes with the strength of the stellar disc bar, as well as evaluating how prolate they become. We are interested in quantifying the shape of the halo bar, not the overall shape of the halo. As the stellar bar grows, the formerly spherical centre of the halo becomes elongated, and this is where we should look for a bar. Following e.g.~\citet{AthanassoulaMisiriotis2002, Machado2010}, we sorted the halo particles by local density and divided them into bins containing the same number of particles. The axis ratios were computed from the eigenvalues of the inertia tensor, using the particles within each ellipsoidal shell of nearly constant density. Thus the minor-to-major ($c/a$) and intermediate-to-major ($b/a$) axis ratios are obtained at each bin. The mean radius of the particles within each bin gives us radial profiles of the axis ratios. The choice of number of particles was such that the average semi-major axis of the innermost density bin is approximately 1\,kpc. In what follows, the $b/a$ and $c/a$ axis ratios refer to this innermost bin.

\begin{figure}
\includegraphics[width=0.6\columnwidth]{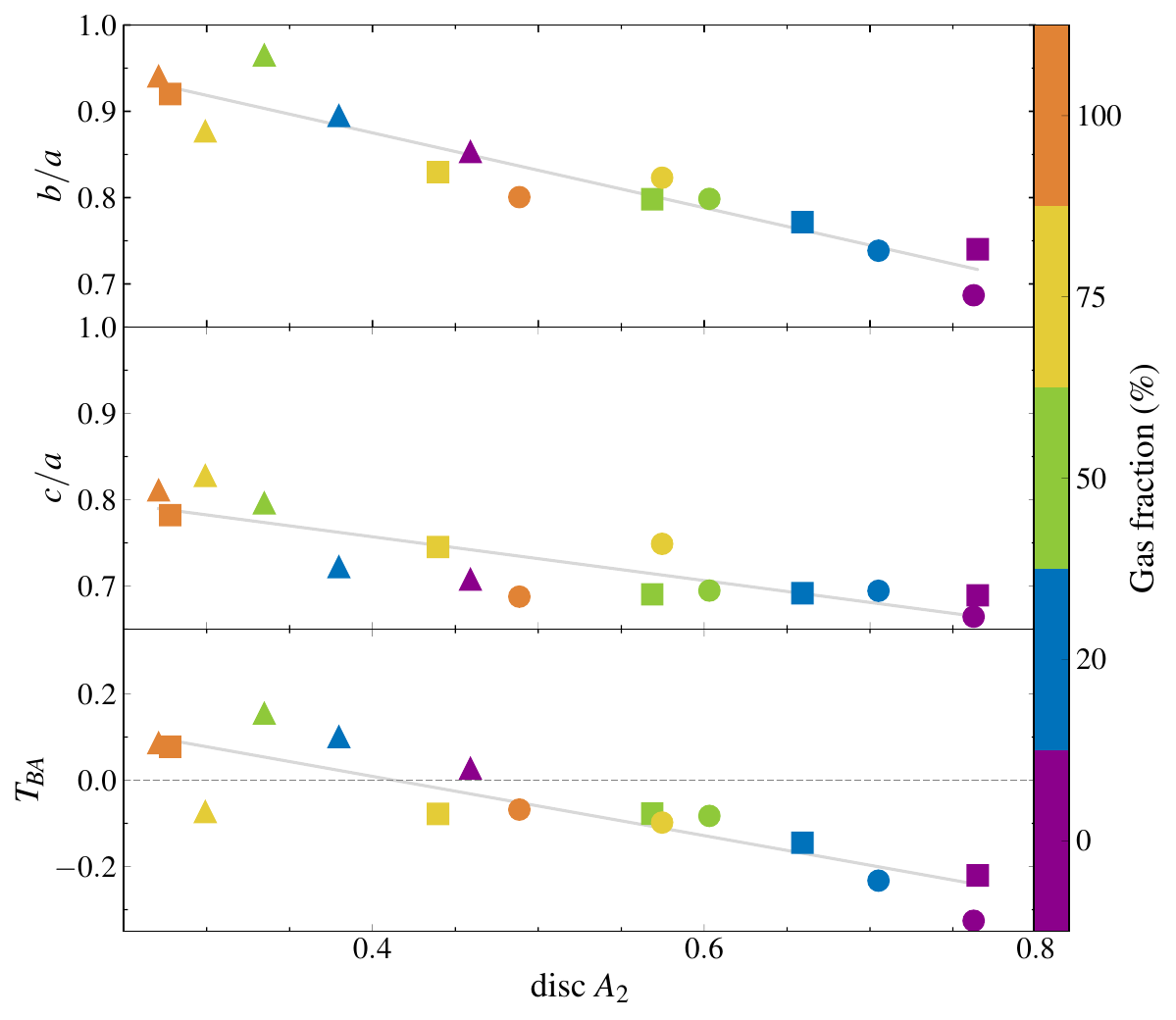}
\caption{From top to bottom, dark matter halo axis ratios ($b/a$ and $c/a$) and triaxiality parameter $T_{BA}$, respectively, as a function of disc bar strength at $t$ = 10\,Gyr. Circles, squares and triangles stand for haloes 1, 2 and 3, respectively. Colour is according to initial gas fraction.}
\label{shape_and_T_vsA2}
\end{figure}

The triaxiality parameter $T_{BA}$, defined as:
\begin{equation}
T_{BA}=\frac{b^2 - c^2}{b^2 + c^2} - \frac{a^2 - b^2}{a^2 + b^2},
\label{boily}
\end{equation}
is one of the possible ways to compute triaxiality. Its advantages reside in its smaller sensitivity to noise and smaller errors if \mbox{$a\approx b\approx c$}~\citep{2006MNRAS.369..608B}. It ranges from $[-1,1]$ (negative for prolate and positive for oblate shapes). Figure~\ref{shape_and_T_vsA2} shows the axial ratios of the dark matter halo ($b/a$ and $c/a$) as well as the triaxiality parameter $T_{BA}$ as a function of the disc $A_2$, for all the 15 runs at $t$ = 10\,Gyr. Circles, squares and triangles stand for haloes 1, 2 and 3, respectively. Colour is according to initial gas fraction.

\begin{figure}[h]
\includegraphics[width=0.5\columnwidth]{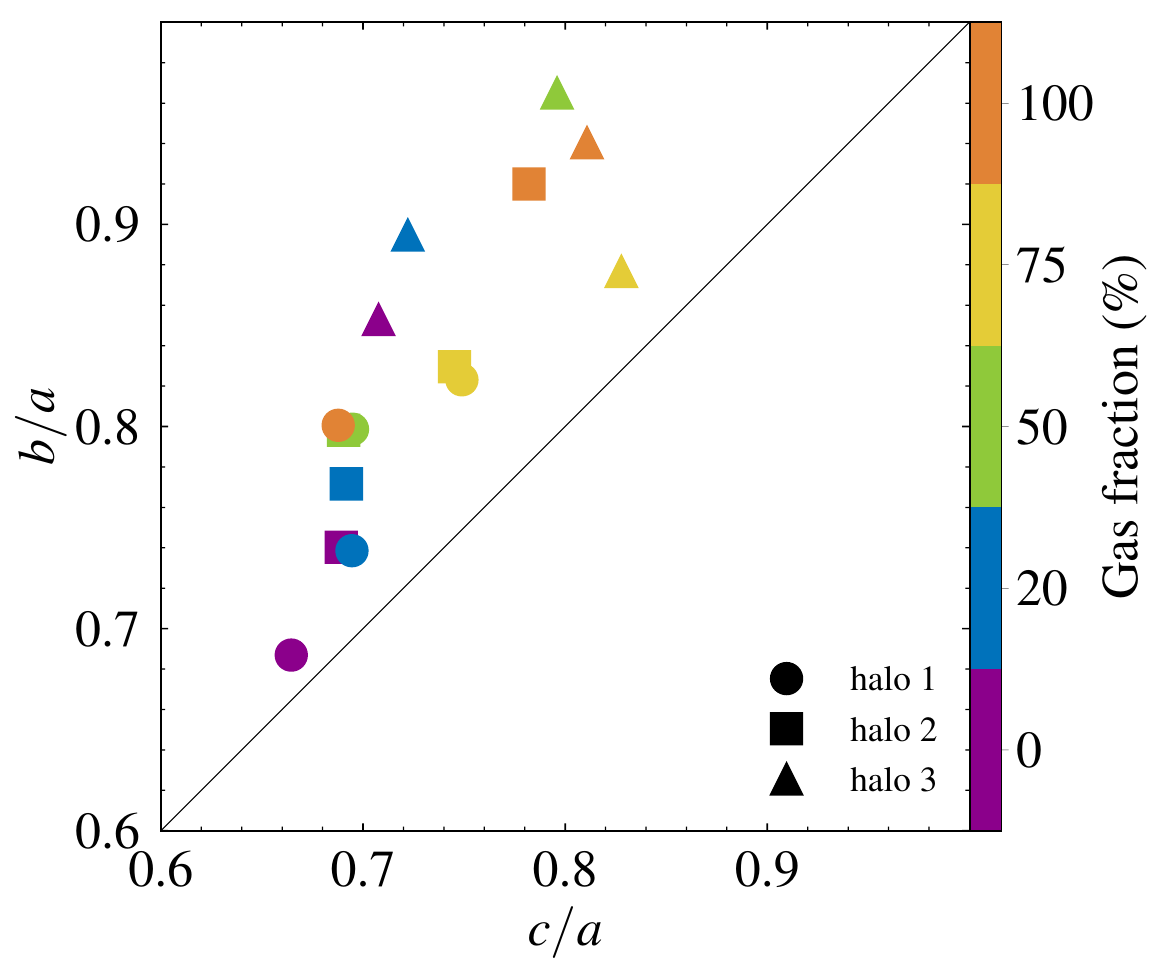}
\caption{Dark matter halo $b/a$ and $c/a$ axis ratios as a function of one another at $t$ = 10\,Gyr. Circles, squares and triangles stand for haloes 1, 2 and 3, respectively. Colour is according to initial gas fraction. Data above close to the diagonal line indicate prolateness of the halo.}
\label{shape_vs_shape}
\end{figure}

Galaxies with strong disc bars host more elongated (small $b/a$ and $c/a$, although this correlation is weaker for the latter) and more prolate ($b/a$ closer to $c/a$) halo bars. The first panel of Fig.~\ref{shape_and_T_vsA2} shows a quite strong correlation between $b/a$ and disc $A_2$ (Spearman $\rho = -0.94$, $p = 2.1\times10^{-7}$). In the second and third panels of Fig.~\ref{shape_and_T_vsA2}, the correlations $c/a$ vs.~$A_2$ and $T_{BA}$ vs.~$A_2$ are nearly as well defined ($\rho = -0.81$, $p = 2.2\times10^{-4}$ and $\rho = -0.88$, $p = 1.6\times10^{-5}$, respectively). Nonetheless, the halo bars in strongly barred galaxies are vertically flatter. In the third panel, we see that most galaxies with halo 3 (triangles) evolve to oblate shapes, regardless of their gas fractions, and two-thirds of the galaxies develop prolate shapes (in special, low gas ones). Thus, the halo bars are mostly prolate in strongly barred galaxies.

In summary, the axial ratios of the inner dark matter halo ($b/a$ and $c/a$) were found to be tighly correlated with the strength of the stellar bar. Galaxies with strong bars host more elongated (small $b/a$ and $c/a$, although this correlation is weaker for the latter) and more prolate ($b/a$ closer to $c/a$) halo bars. In Figure~\ref{shape_vs_shape} we plotted $b/a$ axis ratios as a function of $c/a$. Circles, squares and triangles stand for haloes 1, 2 and 3, respectively. Points close to the diagonal line indicate prolateness of the halo bar within the analysed region. Galaxies with strong disc bars (namely, small initial gas fraction and spherical halo) have halo bars closer to prolateness than the others, with few exceptions. Moreover, models with initially triaxial halos and gas in the discs tend to become oblate at the end of simulations.

\subsection{Inner halo density}
\label{sec:innerdens}

The study of the inner halo density could provide relevant information to the potential direct detection of dark matter. \citet{Petersen2016} found that the density of the halo reacts to the presence of the stellar disc up to a much larger radius than that of the stellar bar. Other studies have estimated the local dark matter enhancement in simulations due to the presence of the disc \citep[e.g][]{Pillepich2014, Bozorgnia2016, Kelso2016, Sloane2016, 2018MNRAS.481.1950L}. Observationally, recent Gaia data has been used to try to constrain the expected halo density of the Milky Way \citep[e.g.][]{2019PhRvD..99b3012E, 2019JCAP...12..013I, 2020PhRvD.101f3026B}. 

\begin{figure}
	\includegraphics[width=0.7\columnwidth]{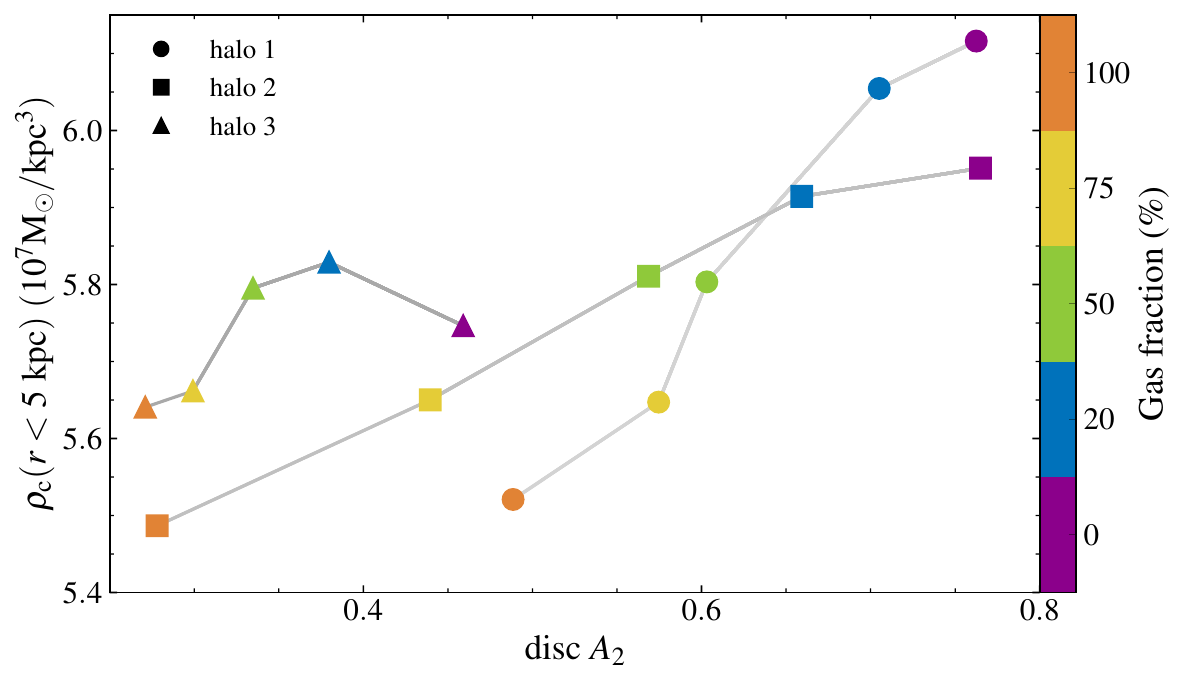} \caption{Central dark matter density (inside 5~kpc) as a function of disc bar strength at $t$ = 10\,Gyr. Circles, squares and triangles stand for haloes 1, 2 and 3, respectively. Colour is according to initial gas fraction. For a better visualisation, grey lines connect galaxies of the same initial halo (halo 1, halo 2 or halo 3)}
    \label{inner_density_vsA2}
\end{figure}

Here we measure the increase of dark matter density in the inner halo at the end of the simulation ($t$ = 10\,Gyr). In Figure~\ref{inner_density_vsA2} we show the average dark matter density within 5\,kpc as a function of bar strength. Stronger bars induce slightly larger inner halo densities. Haloes 1 and 2 show clearer correlations. For the case of halo1, the Pearson correlation coefficient is $r = 0.98$, $p = 0.02$; for halo 2, $r = 0.99$, $p = 0.01$; and for halo3, $r = 0.57$, $p = 0.31$. Thus strongly barred galaxies might be expected to have undergone a small increase in their central dark matter density, compared to weakly barred galaxies.

\subsection{Experiments with nonresponsive discs}
\label{sec:frozen}
The correlations explored so far point to the important role played by the strength of the stellar bar in shaping the properties of the halo bar. In all these fully self-consistent simulations, the disc and halo components are allowed to exchange angular momentum while both bars develop together through this dynamical interplay. But particles in the inner halo would presumably be driven into elongated orbits by the gravitational forces due to a strong stellar bar, even if the disc were non-responsive. In order to disentangle these effects, we devised three numerical experiments aiming to evolve live halo particles in the presence of non-responsive discs. The first one involves a frozen disc (simulation 101-F), the second one employs analytic potentials (simulation 101-A) and, in the third one (simulation 101-R), a rotating rigid disc is used. Galaxy 101 is used as the basis for these experiments because it has the strongest bars. These tests should allow us to evaluate how much of the halo bar growth is due to it being merely driven by the stellar bar. Compared to the self-consistent $N$-body simulations, these experiments are unrealistic by construction. They serve to isolate the effects of interest, by artificially suppressing the responsiveness of the stellar disc.

\subsubsection{Frozen disc}

In the first experiment (101-F) the halo is evolved in the presence of a frozen disc. New initial conditions were constructed in the following manner: the disc particles were taken from the final snapshot ($t=10$\,Gyr) of galaxy 101 and inserted into the initial halo of galaxy 101 (taken from $t=0$). Thus a stellar disc with a strong bar already formed is embedded within an initially spherical halo. This new simulation was then run for 10\,Gyr with a modified version of the code, such that the disc is rigid but its mass does affect the live halo. Halo particles feel their own self-gravity and also feel the gravitational forces due to the frozen disk particles. On the other hand, the disc particles themselves do not undergo gravitational accelerations at all and are kept fixed in place with zero velocity at every time step. The frozen disc is non-rotating.

\begin{figure}
\includegraphics[width=0.7\columnwidth]{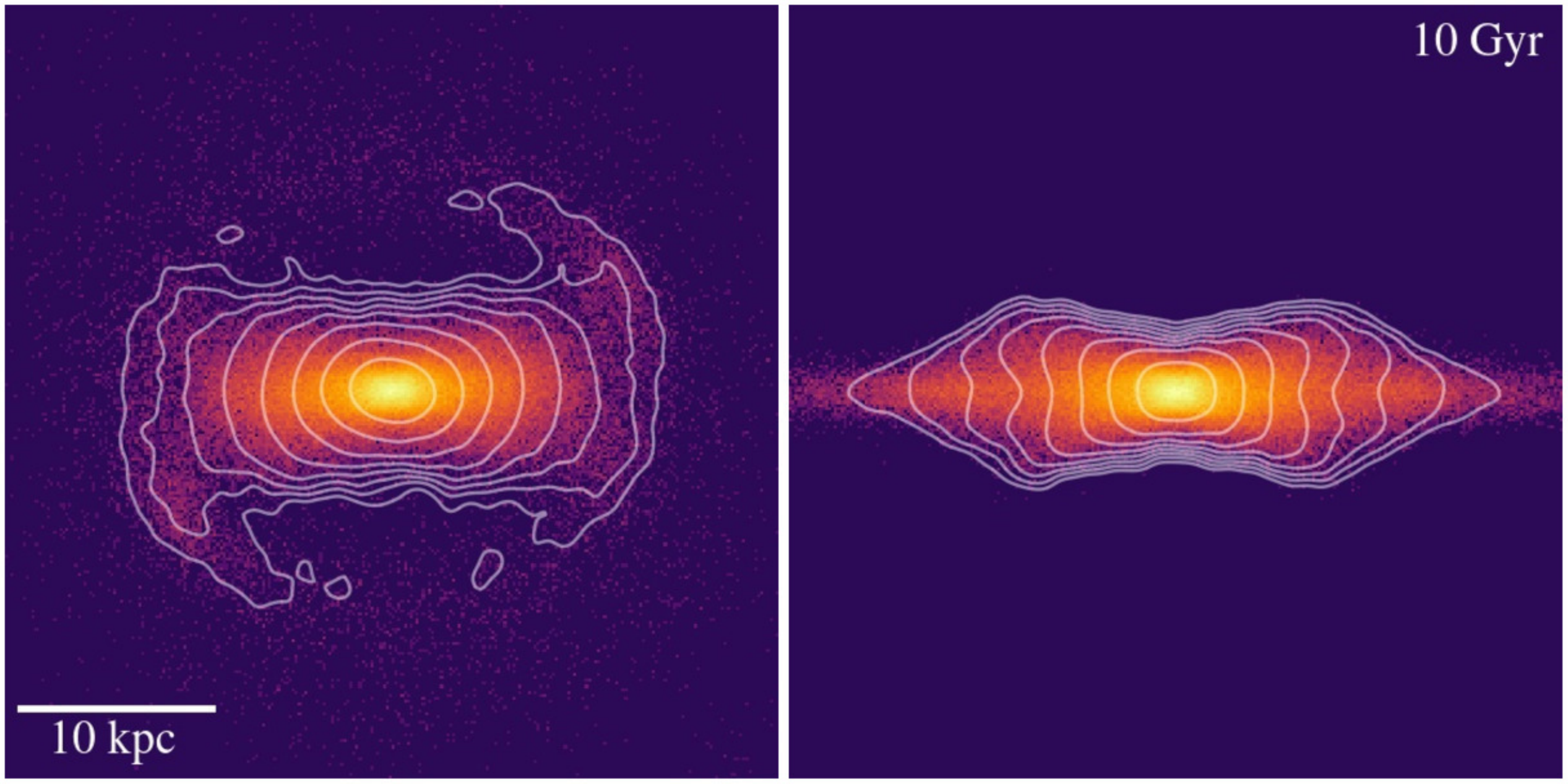}
    \caption{Face-on and side-on projections of the stellar disc from model 101 at $t=10$\,Gyr. Contours represent the potential and colours represent the density. These frozen particles were used in run 101-F.}
    \label{pot}
\end{figure}

For the frozen disc used in simulation 101-F, the gravitational potential is due to mass of the fixed $N$-body particles. Figure~\ref{pot} displays the shape of this potential in a face-on and in a side-on projection. The colours represent the projected stellar mass density, while the contour lines represent the potential, which is generally rounder. The snapshot comes from the end ($t=10$\,Gyr) of the default simulation 101 and the snapshot was rotated so that the bar is aligned with the $x$ axis.

\begin{figure}
\includegraphics[width=0.7\columnwidth]{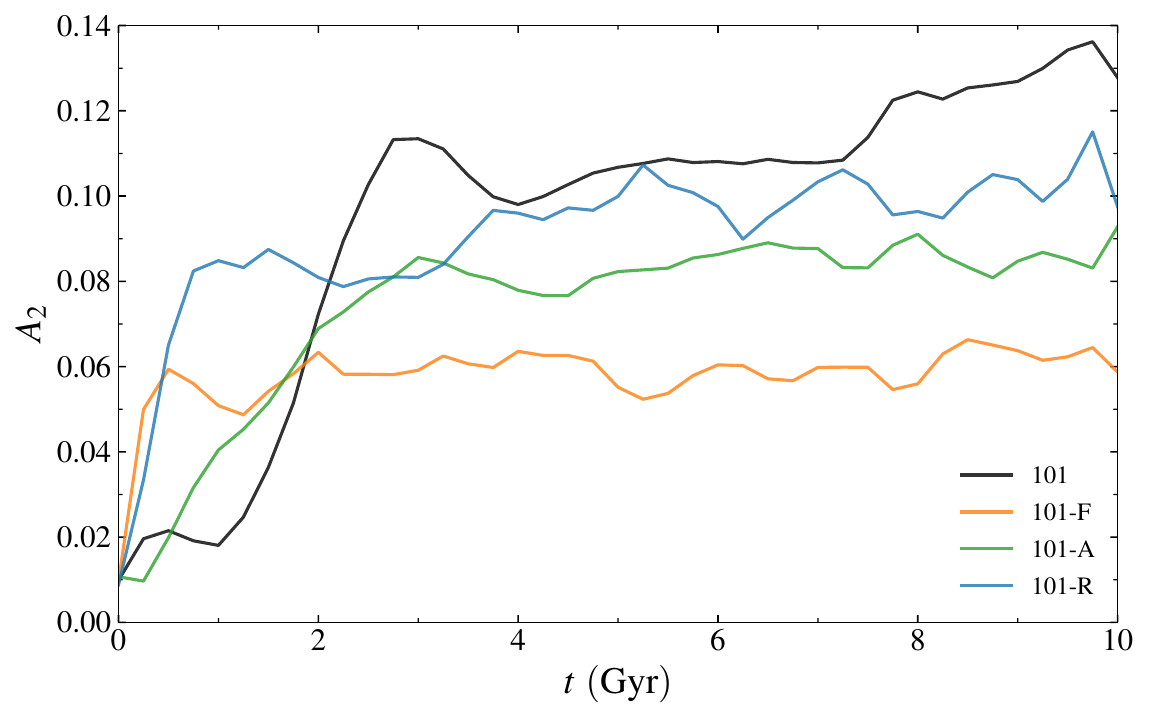}
\caption{Strength of the halo bar of galaxies 101, 101-F and 101-A as a function of time. All galaxies evolve inside live haloes. Galaxy 101 (black line) is the default model and it features a live $N$-body disc. Galaxy 101-F (orange) evolves in the presence of a frozen barred disc potential. Galaxy 101-A (green) has a time-dependent analytic potential emulating the growth of a barred disc. Galaxy 101-R (blue) has a rotating rigid disc.}
\label{frozen}
\end{figure}

We find that simulation 101-F indeed develops a halo bar, albeit weaker than the reference model 101. The resulting evolution of the halo $A_2$ is shown in Figure~\ref{frozen}. This means that a halo bar can form even in the case of a simple, rigid, non-axisymmetric driving. However, its strength is approximately half as much as that of the corresponding self-consistent simulation. We also notice in Figure~\ref{frozen} that the halo bar of 101-F develops earlier; in fact it begins to grow almost immediately after the start of the simulation. This is understandable, since the stellar bar driving this growth is the already strong bar from the end of the original simulation; and this stellar bar is static in 101-F. There is no indication of relevant secular evolution during this 10-Gyr period.

In simulation 101-F, the bar strength is approximately constant over a very extended time span. For this reason, it could be used when one needs cases where it is necessary that the $A_2$ stays constant over a very long time scale \citep[e.g.][]{2016MNRAS.463.3499W}, as long as the early phase of sharp $A_2$ increase is disregarded.

\subsubsection{Analytic potentials}

Frozen particles could in principle be approximated by some rigid analytic potential. However, a time-dependent analytic potential is a suitable way of introducing some temporal evolution -- without bringing back the fully self-consistent setup. In the experiment 101-A, rather than using $N$-body particles, we employ a time-dependent analytic potential to represent the barred disc and evaluate how the live halo particles respond to it. The analytic potential is non-rotating. Galaxy 101 is, by design, quite similar to the default $N$-body model of \citet{Machado2010} (spherical halo with circular disc), which in turn was employed as the basis for the development of a multi-component time-dependent analytic potential that \citet{Manos_2014} and \citet{Machado_2016} used to study chaotic motion in barred galaxies. Therefore, a very convenient time-dependent analytic potential already exists to represent the disc of galaxy 101 within a good approximation.

For the analytic potential employed in simulation 101-A, the galaxy is represented by the sum of two components: a Miyamoto-Nagai disc \citep{1975PASJ...27..533M} plus a Ferrers bar \citep{Fer1877}. It should be noted that, throughout this paper, the term `disc' is meant to be understood as encompassing all of the stars. Within the context of the present subsection, however, it becomes necessary to distinguish between two stellar components: the stars belonging to the bar (represented by the Ferrers potential); and the remainder of the stars (represented by the Miyamoto-Nagai potential). The Miyamoto-Nagai potential is axisymmetric and its parameters control the disc mass, vertical thickness and radial scale length. The parameters of the Ferrers bar control the bar mass and ellipsoidal shape. The disc mass decreases at the expense of the growing bar mass, such that the stellar mass ($5\times10^{10}\,{\rm M_{\odot}}$) remains constant. Fits were performed for each of those parameters at each snapshot; finally smooth polynomials were fitted, such that the combined potentials could be written as a function of time. \citet{Manos_2014} showed that the resulting model -- although necessarily an idealised simplification -- produced circular velocity curves that approximate the original $N$-body simulation to a very acceptable degree. This indicates that the analytic model must be capturing at least the global dynamical features of the system. Now, taking again the initially spherical live halo of model 101, we ran simulation 101-A for 10\,Gyr with a modified version of \textsc{gadget2}, onto which we grafted the time-dependent analytic potential. At each simulation time step, the gravitational forces due to the Ferrers plus Miyamoto-Nagai potentials are added. The details of the potential are given in \citet{Manos_2014} and summarised in what follows.

The Ferrers bar is a triaxial ellipsoid whose density is given by:
\begin{eqnarray}
\rho(x,y,z) = \left\{ \begin{array}{l l}
\rho_{\rm c}(1-m^{2})^{2}& \quad \textrm{if} \quad m<1,\\
\quad 0& \quad \textrm{if} \quad m \geq 1,\\
\end{array} \right.
\end{eqnarray}
where the central density is $\rho_{\rm c}=\frac{105}{32\pi}\frac{M_{B}(t)}{abc}$; $G$ is the gravitational constant, $m^{2}=\frac{x^{2}}{a^{2}}+\frac{y^{2}}{b^{2}}+\frac{z^{2}}{c^{2}}$,
\mbox{$a>b>c> 0$}, and $a,b$ and $c$ are the semi-axes of the ellipsoid. The mass of the bar is $M_{B}(t)$ and it evolves in time. The resulting potential of the bar is then:
\begin{equation}\label{Ferr_pot}
V_{B}(t)= -\pi Gabc \frac{\rho_{\rm c}}{3}\int_{\lambda}^{\infty} \frac{du}{\Delta (u)} (1-m^{2}(u))^{3},
\end{equation}
where $m^{2}(u)=\frac{x^{2}}{a^{2}+u}+\frac{y^{2}}{b^{2}+u}+\frac{z^{2}}{c^{2}+u}$, and
$\Delta^{2} (u)=({a^{2}+u})({b^{2}+u})({c^{2}+u})$; $\lambda$ is the unique
positive solution of $m^{2}(\lambda)=1$, outside the bar ($m \geq 1$), whereas
$\lambda=0$ inside the bar. The analytical formulae for the corresponding forces are provided in \citet{Pfenniger1984}. Those were the forces -- due to the bar -- we implemented in the modified version of \textsc{gadget2}. In the time-dependent model of \citet{Manos_2014}, not only is the mass of the bar a function of time, but so are the parameters $a(t),b(t)$ and $c(t)$. During the 10\,Gyr of the simulation, the mass of the bar grows smoothly from nearly zero to approximately $3\times10^{10}\,{\rm M_{\odot}}$. By the end of the simulation, the shape parameters have reached about 8, 3 and 2\,kpc approximately.

The disc potential is represented by the Miyamoto-Nagai potential:
 \begin{equation}\label{eq:MNPot}
   V_{D}(t)=- \frac{GM_{D}(t)}{\sqrt{x^{2}+y^{2}+(A+\sqrt{z^{2}+B^{2}})^{2}}},
\end{equation}
where $A$ and $B$ are respectively the horizontal and the vertical scale lengths. The mass of the disc is $M_{D}(t)$ and it is understood as the stellar mass not including the mass of the bar; thus it decreases in time. The vertical scale length of the disc grows from nearly zero to about 0.5\,kpc, while the radial scale length evolves from 2.5 to 0.7\,kpc approximately. Likewise, the corresponding forces due to the disc were also included in the modification of \textsc{gadget2}.

The halo bar also forms in the presence of this analytic potential. The strength of the halo bar shown in Figure~\ref{frozen} indicates that this halo bar is somewhat stronger than in the case of the frozen disc, but still not as strong as the reference $N$-body model. Model 101-A is somewhat less artificial than 101-F at least in the sense that the analytic bar grows smoothly in time. (Nevertheless the Ferrers potential still gives an idealised stellar bar, whose ellipsoidal shape cannot reproduce the complexity of the $N$-body bar and whose mass may be slightly overestimated.) The early growth of the halo bar in 101-A is consequently not as abrupt as in 101-F, but still differs from the reference model. In the early growth as well as in the final strength, model 101-A may be regarded as an intermediate case between 101 and 101-F. Still the conclusion would be that the live disc particles in the fully self-consistent simulation play a role that cannot be emulated by a non-responsive (even if non-fixed) analytic disc potential.

\subsubsection{Rigid disc}

\begin{figure}
\includegraphics[width=0.7\columnwidth]{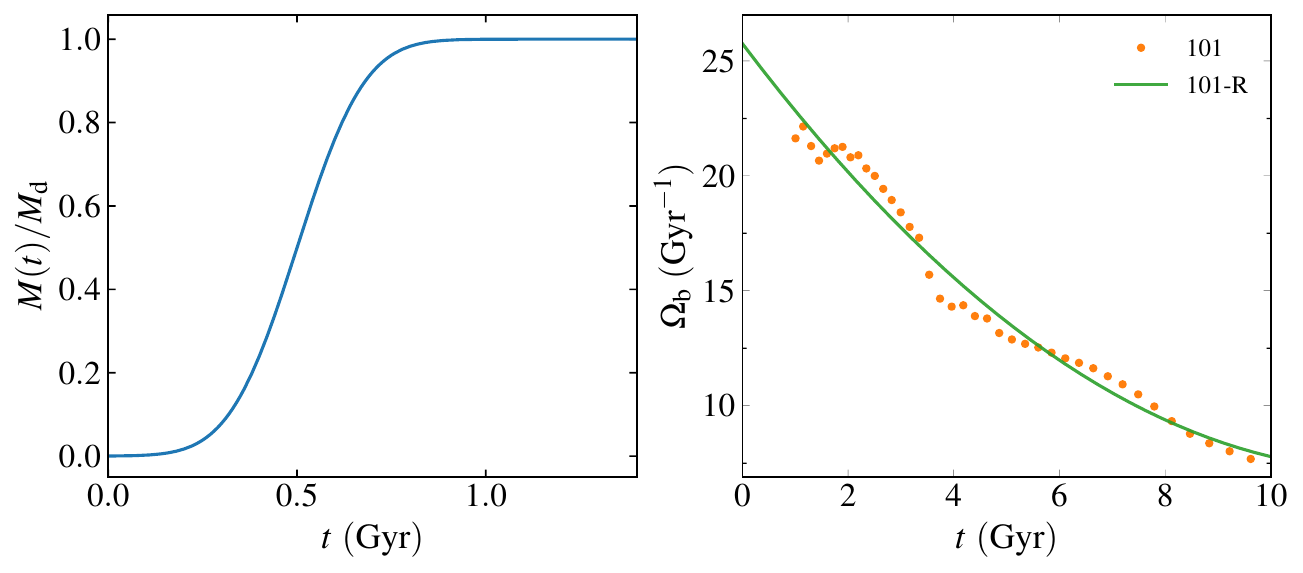}
    \caption{Left: in the rotating rigid disc 101-R, the stellar mass is grown smoothly from 0 to $M_{\rm d}$ during the first 1\,Gyr. Right: the measured pattern speed from the bar in model 101 (symbols) was used to impose a rotation to model 101-R (line).}
    \label{rot}
\end{figure}

Finally, the third experiment includes rotation, unlike the two previous experiments. In simulation 101-R, the disc particles (meaning all of the stars) taken from 101 at $t=10$\,Gyr are unresponsive, as is also the case of 101-F. However, now, the entire disc is made to rotate as a rigid body. This is done by imposing the pattern speed $\Omega_{\rm b}(t)$ that was measured from model 101. Figure~\ref{rot} shows the measured pattern speed and the fitted polynomial that was used to impose a smooth rotation on the disc of 101-R. At $t=0$\,Gyr, the mass of the disc of 101-R was set to zero. Additionally, the disc had its mass `adiabatically' grown from 0 until $M_{\rm d}=5\times10^{10}\,{\rm M_{\odot}}$ during the first 1 Gyr, according to a smooth function shown in Figure~\ref{rot}:
\begin{equation}
M(t) = \frac{M_{\rm d}}{2} \left[ 1 + {\rm erf}\left( \frac{t-\mu}{\sqrt{2\sigma^2}} \right) \right], 
\end{equation}
with parameters $\mu=0.5$\,Gyr and $\sigma=0.14$\,Gyr. The code was modified to alter the masses of the disc particles at each time step of the computations. Also at each time step, the positions of all disc particles were turned by a small angle, such that this rigid bar effectively rotates with the imposed pattern speed.

The result of the rotating rigid disc of 101-R is to induce a halo bar which is stronger than 101-F and 101-A (Figure~\ref{frozen}), but still not quite as strong as the live case 101. The comparison between 101-R and 101-F highlights the difference between a rotating and a non-rotating rigid bar. This comparison indicates that a rotating stellar bar is more effective at driving the inner halo into an elongated shape. Even though experiment 101-R is quite artificial in its setup, it is the one that most closely approaches the halo bar strength of 101. It should be noted in the $A_2(t)$ curves of Figure~\ref{frozen} that the halo bar strength of the fully self-consistent simulation overtakes all the simplified simulations at approximately the same moment ($\sim$2\,Gyr).

These experiments were designed merely to isolate a given mechanism and explore its numerical consequences in a theoretical frame. In this sense, each experiment on its own was not meant to produce a realistic evolution of the bar. Nor was the goal of each experiment to approach the $N$-body model as closely as possible. Additionally their conclusions apply to the particular case of one strongly barred galaxy. Nevertheless, these results indicate that the formation of the halo bar may be attributed in part -- but not entirely -- to the halo particles being driven by the potential of the stellar bar.

\section{Discussion and conclusions}

Cosmological simulations have indicated that dark matter haloes tend to be prolate or generally triaxial. However, the presence of baryonic discs acts to circularise the inner halo. If a strong stellar bar develops, the innermost region of the halo may again be driven into an elongated shape.
In order to investigate this interplay, we have systematically quantified morphological properties of the stellar bar and of the dark matter bar in a set of simulations from \cite{Athanassoula2013}. The initial conditions of those simulations had been prepared to provide a meticulous exploration of the parameter space of halo shapes and disc gas fractions. Moreover, those are hydrodynamical simulations including gas and also the sub-grid physics of star formation, cooling and feedback. The resulting models are well suited for our purposes because they cover a wide range of bar strengths, thus sampling different regimes from strongly barred to nearly no bar.

Measuring the strengths of the halo bar, we have shown that it generally follows the trends of the stellar bar, provided the latter is sufficiently strong. Although the halo bar is weaker than the stellar bar by an order of magnitude, they seem to form at the same time and their evolutions run in parallel for the strongest cases. For the weaker cases, the halo bar evolution is not as steep as the stellar bar. If the stellar bar is very weak, the halo bar is nearly undetectable. Nevertheless a tight correlation was found between the strengths of the bars, indicating that the strength of the halo bar is governed by the stellar bar.

We have analysed the shape of the inner halo and found that the $b/a$ axis ratio is strongly correlated with the stellar bar strength, meaning that the halo bars are more elongated in the strongly barred cases. Stronger halo bars also tend to be vertically flatter, but the correlation with $c/a$ is mild. This translates into a more prolate, bar-like shape of the halo bar in the strong regime. 

We have also looked into how the bar can enhance the central dark matter density. We found that stronger bars induce slightly larger inner halo densities, in comparison with other similar weakly-barred galaxies. The expectations of dark matter densities in the centres of galaxies could have implications for the future search for direct detection, and the increment due to the halo bars might be relevant \citep{2005PhRvD..72h3503A, 2010JCAP...02..012L, Petersen2016D, Petersen2016}.

Finally, we performed numerical experiments designed to evaluate the direct role of the stellar bar in driving the formation of the halo bar. In one case, a strongly barred frozen disc was capable of inducing a bar in a live responsive halo, but this halo bar was weaker by a factor of two compared to the fully self-consistent simulation. Similarly, a time-dependent analytic bar potential induces a halo bar as well, but not to the same degree as the live disc case. A rotating rigid bar still did not reach the same halo bar strength as the fully self-consistent simulation, but it did produce the strongest halo bar among the artificial experiments, indicating the importance of rotation. These findings suggest that the formation of the halo bar can be attributed only in part to the mere forcing by the stellar bar potential, and that the presence of a live responsive disc indeed contributes to the dynamical mechanisms that produce the halo bar.

\vspace{6pt}

\authorcontributions{Conceptualization, RM, EA; methodology, DM, RM, TM; software, DM, RM, TM; formal analysis, DM; resources, RM, EA; data curation, RM, EA; writing---original draft preparation, DM; writing---review and editing, RM, EA; visualization, DM; supervision, RM, EA; project administration, RM, EA; funding acquisition, RM, EA. All authors have read and agreed to the published version of the manuscript.}

\funding{DM acknowledges support from CNPq and UTFPR. RM acknowledges support from the Brazilian agency \textit{Conselho Nacional de Desenvolvimento Cient\'ifico e Tecnol\'ogico} (CNPq) through grants  406908/2018-4, 303426/2018-7, and 307205/2021-5 and from \textit{Funda\c c\~ao de Apoio \`a Ci\^encia, Tecnologia e Inova\c c\~ao do Paran\'a} through grant 18.148.096-3 -- NAPI \textit{Fen\^omenos Extremos do Universo}. EA acknowledges financial support from the CNES (\textit{Centre national d'etudes spatiales}, France)}

\dataavailability{The only three new simulations presented here are the response simulations of section 3.4. These can be shared upon reasonable request to the corresponding author.}

\acknowledgments{The authors acknowledge the National Laboratory for Scientific Computing (LNCC/MCTI, Brazil) for providing HPC resources of the SDumont supercomputer, which have contributed to the research results reported within this paper.}

\conflictsofinterest{The authors declare no conflicts of interest. The funders had no role in the design of the study; in the collection, analyses, or interpretation of data; in the writing of the manuscript; or in the decision to publish the results.}

\begin{adjustwidth}{-\extralength}{0cm}

\reftitle{References}
\bibliography{paper}

\begin{thebibliography}{999}

\bibitem[{Eskridge} et~al.(2000){Eskridge}, {Frogel}, {Pogge}, {Quillen},
  {Davies}, {DePoy}, {Houdashelt}, {Kuchinski}, {Ram{\'\i}rez}, {Sellgren},
  {Terndrup}, and {Tiede}]{2000AJ....119..536E}
{Eskridge}, P.B.; {Frogel}, J.A.; {Pogge}, R.W.; {Quillen}, A.C.; {Davies},
  R.L.; {DePoy}, D.L.; {Houdashelt}, M.L.; {Kuchinski}, L.E.; {Ram{\'\i}rez},
  S.V.; {Sellgren}, K.;  et~al.
\newblock {The Frequency of Barred Spiral Galaxies in the Near-Infrared}.
\newblock {\em \aj} {\bf 2000}, {\em 119},~536--544,
  \href{http://arxiv.org/abs/astro-ph/9910479}{{\normalfont
  [arXiv:astro-ph/astro-ph/9910479]}}.
\newblock {\url{https://doi.org/10.1086/301203}}.

\bibitem[{Sheth} et~al.(2008){Sheth}, {Elmegreen}, {Elmegreen}, {Capak},
  {Abraham}, {Athanassoula}, {Ellis}, {Mobasher}, {Salvato}, {Schinnerer},
  {Scoville}, {Spalsbury}, {Strubbe}, {Carollo}, {Rich}, and
  {West}]{Sheth_2008}
{Sheth}, K.; {Elmegreen}, D.M.; {Elmegreen}, B.G.; {Capak}, P.; {Abraham},
  R.G.; {Athanassoula}, E.; {Ellis}, R.S.; {Mobasher}, B.; {Salvato}, M.;
  {Schinnerer}, E.;  et~al.
\newblock {Evolution of the Bar Fraction in COSMOS: Quantifying the Assembly of
  the Hubble Sequence}.
\newblock {\em \apj} {\bf 2008}, {\em 675},~1141--1155,
  \href{http://arxiv.org/abs/0710.4552}{{\normalfont
  [arXiv:astro-ph/0710.4552]}}.
\newblock {\url{https://doi.org/10.1086/524980}}.

\bibitem[{Nair} and {Abraham}(2010)]{2010ApJ...714L.260N}
{Nair}, P.B.; {Abraham}, R.G.
\newblock {On the Fraction of Barred Spiral Galaxies}.
\newblock {\em \apjl} {\bf 2010}, {\em 714},~L260--L264,
  \href{http://arxiv.org/abs/1004.0684}{{\normalfont
  [arXiv:astro-ph.CO/1004.0684]}}.
\newblock {\url{https://doi.org/10.1088/2041-8205/714/2/L260}}.

\bibitem[{Garc{\'\i}a-Barreto}(2011)]{2011RMxAC..40..120G}
{Garc{\'\i}a-Barreto}, J.A.
\newblock {Fraction of Strong Barred Galaxies (SB) in the Nearby Universe, 0
  {\ensuremath{\leq}} z {\ensuremath{\leq}} 0.066 as a function of redshift}.
\newblock In Proceedings of the Revista Mexicana de Astronomia y Astrofisica
  Conference Series,  2011, Vol.~40, {\em Revista Mexicana de Astronomia y
  Astrofisica Conference Series}, pp. 120--120,
  \href{http://arxiv.org/abs/1012.3669}{{\normalfont
  [arXiv:astro-ph.GA/1012.3669]}}.
\newblock {\url{https://doi.org/10.48550/arXiv.1012.3669}}.

\bibitem[{Kraljic} et~al.(2012){Kraljic}, {Bournaud}, and
  {Martig}]{2012ApJ...757...60K}
{Kraljic}, K.; {Bournaud}, F.; {Martig}, M.
\newblock {The Two-phase Formation History of Spiral Galaxies Traced by the
  Cosmic Evolution of the Bar Fraction}.
\newblock {\em \apj} {\bf 2012}, {\em 757},~60,
  \href{http://arxiv.org/abs/1207.0351}{{\normalfont
  [arXiv:astro-ph.GA/1207.0351]}}.
\newblock {\url{https://doi.org/10.1088/0004-637X/757/1/60}}.

\bibitem[{Lee} et~al.(2019){Lee}, {Ann}, and {Park}]{2019ApJ...872...97L}
{Lee}, Y.H.; {Ann}, H.B.; {Park}, M.G.
\newblock {Bar Fraction in Early- and Late-type Spirals}.
\newblock {\em \apj} {\bf 2019}, {\em 872},~97,
  \href{http://arxiv.org/abs/1901.05183}{{\normalfont
  [arXiv:astro-ph.GA/1901.05183]}}.
\newblock {\url{https://doi.org/10.3847/1538-4357/ab0024}}.

\bibitem[{Kormendy}(1979)]{1979ApJ...227..714K}
{Kormendy}, J.
\newblock {A morphological survey of bar, lens, and ring components in
  galaxies: secular evolution in galaxy structure.}
\newblock {\em \apj} {\bf 1979}, {\em 227},~714--728.
\newblock {\url{https://doi.org/10.1086/156782}}.

\bibitem[{Buta} et~al.(2009){Buta}, {Knapen}, {Elmegreen}, {Salo},
  {Laurikainen}, {Elmegreen}, {Puerari}, and {Block}]{2009AJ....137.4487B}
{Buta}, R.J.; {Knapen}, J.H.; {Elmegreen}, B.G.; {Salo}, H.; {Laurikainen}, E.;
  {Elmegreen}, D.M.; {Puerari}, I.; {Block}, D.L.
\newblock {Do Bars Drive Spiral Density Waves?}
\newblock {\em \aj} {\bf 2009}, {\em 137},~4487--4516,
  \href{http://arxiv.org/abs/0903.2008}{{\normalfont
  [arXiv:astro-ph.CO/0903.2008]}}.
\newblock {\url{https://doi.org/10.1088/0004-6256/137/5/4487}}.

\bibitem[{Wang} et~al.(2012){Wang}, {Kauffmann}, {Overzier}, {Tacconi}, {Kong},
  {Saintonge}, {Catinella}, {Schiminovich}, {Moran}, and
  {Johnson}]{2012MNRAS.423.3486W}
{Wang}, J.; {Kauffmann}, G.; {Overzier}, R.; {Tacconi}, L.J.; {Kong}, X.;
  {Saintonge}, A.; {Catinella}, B.; {Schiminovich}, D.; {Moran}, S.M.;
  {Johnson}, B.
\newblock {Quantifying the role of bars in the build-up of central mass
  concentrations in disc galaxies}.
\newblock {\em \mnras} {\bf 2012}, {\em 423},~3486--3501,
  \href{http://arxiv.org/abs/1205.0932}{{\normalfont
  [arXiv:astro-ph.CO/1205.0932]}}.
\newblock {\url{https://doi.org/10.1111/j.1365-2966.2012.21147.x}}.

\bibitem[{Ciambur} et~al.(2017){Ciambur}, {Graham}, and
  {Bland-Hawthorn}]{Ciambur2017}
{Ciambur}, B.C.; {Graham}, A.W.; {Bland-Hawthorn}, J.
\newblock {Quantifying the (X/peanut)-shaped structure of the Milky Way - new
  constraints on the bar geometry}.
\newblock {\em \mnras} {\bf 2017}, {\em 471},~3988--4004,
  \href{http://arxiv.org/abs/1706.09902}{{\normalfont
  [arXiv:astro-ph.GA/1706.09902]}}.
\newblock {\url{https://doi.org/10.1093/mnras/stx1823}}.

\bibitem[{M{\'e}ndez-Abreu} et~al.(2018){M{\'e}ndez-Abreu}, {Costantin},
  {Aguerri}, {de Lorenzo-C{\'a}ceres}, and {Corsini}]{Mendez2018}
{M{\'e}ndez-Abreu}, J.; {Costantin}, L.; {Aguerri}, J.A.L.; {de
  Lorenzo-C{\'a}ceres}, A.; {Corsini}, E.M.
\newblock {The intrinsic three-dimensional shape of galactic bars}.
\newblock {\em \mnras} {\bf 2018}, {\em 479},~4172--4186,
  \href{http://arxiv.org/abs/1805.09481}{{\normalfont
  [arXiv:astro-ph.GA/1805.09481]}}.
\newblock {\url{https://doi.org/10.1093/mnras/sty1694}}.

\bibitem[{Nitschai} et~al.(2020){Nitschai}, {Cappellari}, and
  {Neumayer}]{Nitschai2019}
{Nitschai}, M.S.; {Cappellari}, M.; {Neumayer}, N.
\newblock {First Gaia dynamical model of the Milky Way disc with six phase
  space coordinates: a test for galaxy dynamics}.
\newblock {\em \mnras} {\bf 2020}, {\em 494},~6001--6011,
  \href{http://arxiv.org/abs/1909.05269}{{\normalfont
  [arXiv:astro-ph.GA/1909.05269]}}.
\newblock {\url{https://doi.org/10.1093/mnras/staa1128}}.

\bibitem[{Gadotti}(2011)]{2011MNRAS.415.3308G}
{Gadotti}, D.A.
\newblock {Secular evolution and structural properties of stellar bars in
  galaxies}.
\newblock {\em \mnras} {\bf 2011}, {\em 415},~3308--3318,
  \href{http://arxiv.org/abs/1003.1719}{{\normalfont
  [arXiv:astro-ph.CO/1003.1719]}}.
\newblock {\url{https://doi.org/10.1111/j.1365-2966.2011.18945.x}}.

\bibitem[{Cuomo} et~al.(2020){Cuomo}, {Aguerri}, {Corsini}, and
  {Debattista}]{2020A&A...641A.111C}
{Cuomo}, V.; {Aguerri}, J.A.L.; {Corsini}, E.M.; {Debattista}, V.P.
\newblock {Relations among structural parameters in barred galaxies with a
  direct measurement of bar pattern speed}.
\newblock {\em \aap} {\bf 2020}, {\em 641},~A111,
  \href{http://arxiv.org/abs/2003.07455}{{\normalfont
  [arXiv:astro-ph.GA/2003.07455]}}.
\newblock {\url{https://doi.org/10.1051/0004-6361/202037945}}.

\bibitem[{Ellison} et~al.(2011){Ellison}, {Nair}, {Patton}, {Scudder},
  {Mendel}, and {Simard}]{2011MNRAS.416.2182E}
{Ellison}, S.L.; {Nair}, P.; {Patton}, D.R.; {Scudder}, J.M.; {Mendel}, J.T.;
  {Simard}, L.
\newblock {The impact of gas inflows on star formation rates and metallicities
  in barred galaxies}.
\newblock {\em \mnras} {\bf 2011}, {\em 416},~2182--2192,
  \href{http://arxiv.org/abs/1106.1177}{{\normalfont
  [arXiv:astro-ph.CO/1106.1177]}}.
\newblock {\url{https://doi.org/10.1111/j.1365-2966.2011.19195.x}}.

\bibitem[{Masters} et~al.(2012){Masters}, {Nichol}, {Haynes}, {Keel},
  {Lintott}, {Simmons}, {Skibba}, {Bamford}, {Giovanelli}, and
  {Schawinski}]{2012MNRAS.424.2180M}
{Masters}, K.L.; {Nichol}, R.C.; {Haynes}, M.P.; {Keel}, W.C.; {Lintott}, C.;
  {Simmons}, B.; {Skibba}, R.; {Bamford}, S.; {Giovanelli}, R.; {Schawinski},
  K.
\newblock {Galaxy Zoo and ALFALFA: atomic gas and the regulation of star
  formation in barred disc galaxies}.
\newblock {\em \mnras} {\bf 2012}, {\em 424},~2180--2192,
  \href{http://arxiv.org/abs/1205.5271}{{\normalfont
  [arXiv:astro-ph.CO/1205.5271]}}.
\newblock {\url{https://doi.org/10.1111/j.1365-2966.2012.21377.x}}.

\bibitem[{Cheung} et~al.(2013){Cheung}, {Athanassoula}, {Masters}, {Nichol},
  {Bosma}, {Bell}, {Faber}, {Koo}, {Lintott}, {Melvin}, {Schawinski}, {Skibba},
  and {Willett}]{2013ApJ...779..162C}
{Cheung}, E.; {Athanassoula}, E.; {Masters}, K.L.; {Nichol}, R.C.; {Bosma}, A.;
  {Bell}, E.F.; {Faber}, S.M.; {Koo}, D.C.; {Lintott}, C.; {Melvin}, T.;
  et~al.
\newblock {Galaxy Zoo: Observing Secular Evolution through Bars}.
\newblock {\em \apj} {\bf 2013}, {\em 779},~162,
  \href{http://arxiv.org/abs/1310.2941}{{\normalfont
  [arXiv:astro-ph.CO/1310.2941]}}.
\newblock {\url{https://doi.org/10.1088/0004-637X/779/2/162}}.

\bibitem[{Williams} et~al.(2012){Williams}, {Bureau}, and
  {Kuntschner}]{Williams_2012}
{Williams}, M.J.; {Bureau}, M.; {Kuntschner}, H.
\newblock {Secular evolution in action: central values and radial trends in the
  stellar populations of boxy bulges}.
\newblock {\em \mnras} {\bf 2012}, {\em 427},~L99--L103,
  \href{http://arxiv.org/abs/1209.3167}{{\normalfont
  [arXiv:astro-ph.CO/1209.3167]}}.
\newblock {\url{https://doi.org/10.1111/j.1745-3933.2012.01353.x}}.

\bibitem[{Chown} et~al.(2019){Chown}, {Li}, {Athanassoula}, {Li}, {Wilson},
  {Lin}, {Mo}, {Parker}, and {Xiao}]{2019MNRAS.484.5192C}
{Chown}, R.; {Li}, C.; {Athanassoula}, E.; {Li}, N.; {Wilson}, C.D.; {Lin}, L.;
  {Mo}, H.; {Parker}, L.C.; {Xiao}, T.
\newblock {Linking bar- and interaction-driven molecular gas concentration with
  centrally enhanced star formation in EDGE-CALIFA galaxies}.
\newblock {\em \mnras} {\bf 2019}, {\em 484},~5192--5211,
  \href{http://arxiv.org/abs/1810.08624}{{\normalfont
  [arXiv:astro-ph.GA/1810.08624]}}.
\newblock {\url{https://doi.org/10.1093/mnras/stz349}}.

\bibitem[{Laurikainen} et~al.(2004){Laurikainen}, {Salo}, and
  {Buta}]{2004ApJ...607..103L}
{Laurikainen}, E.; {Salo}, H.; {Buta}, R.
\newblock {Comparison of Bar Strengths and Fractions of Bars in Active and
  Nonactive Galaxies}.
\newblock {\em \apj} {\bf 2004}, {\em 607},~103--124,
  \href{http://arxiv.org/abs/astro-ph/0111376}{{\normalfont
  [arXiv:astro-ph/astro-ph/0111376]}}.
\newblock {\url{https://doi.org/10.1086/383462}}.

\bibitem[{Athanassoula}(1996)]{1996ASPC...91..309A}
{Athanassoula}, E.
\newblock {Evolution of Bars in Isolated and in Interacting Disk Galaxies}.
\newblock In Proceedings of the IAU Colloq. 157: Barred Galaxies; {Buta}, R.;
  {Crocker}, D.A.; {Elmegreen}, B.G., Eds.,  1996, Vol.~91, {\em Astronomical
  Society of the Pacific Conference Series}, p. 309.

\bibitem[{Debattista} and {Sellwood}(1996)]{1996ASPC...91..357D}
{Debattista}, V.P.; {Sellwood}, J.A.
\newblock {Dynamical Friction in Barred Galaxies}.
\newblock In Proceedings of the IAU Colloq. 157: Barred Galaxies; {Buta}, R.;
  {Crocker}, D.A.; {Elmegreen}, B.G., Eds.,  1996, Vol.~91, {\em Astronomical
  Society of the Pacific Conference Series}, p. 357.

\bibitem[{Debattista} and {Sellwood}(1998)]{1998ApJ...493L...5D}
{Debattista}, V.P.; {Sellwood}, J.A.
\newblock {Dynamical Friction and the Distribution of Dark Matter in Barred
  Galaxies}.
\newblock {\em \apjl} {\bf 1998}, {\em 493},~L5--L8,
  \href{http://arxiv.org/abs/astro-ph/9710039}{{\normalfont
  [arXiv:astro-ph/astro-ph/9710039]}}.
\newblock {\url{https://doi.org/10.1086/311118}}.

\bibitem[{Berentzen} et~al.(1998){Berentzen}, {Heller}, {Shlosman}, and
  {Fricke}]{1998MNRAS.300...49B}
{Berentzen}, I.; {Heller}, C.H.; {Shlosman}, I.; {Fricke}, K.J.
\newblock {Gas-driven evolution of stellar orbits in barred galaxies}.
\newblock {\em \mnras} {\bf 1998}, {\em 300},~49--63,
  \href{http://arxiv.org/abs/astro-ph/9806138}{{\normalfont
  [arXiv:astro-ph/astro-ph/9806138]}}.
\newblock {\url{https://doi.org/10.1046/j.1365-8711.1998.01836.x}}.

\bibitem[{Debattista} and {Sellwood}(2000)]{2000ApJ...543..704D}
{Debattista}, V.P.; {Sellwood}, J.A.
\newblock {Constraints from Dynamical Friction on the Dark Matter Content of
  Barred Galaxies}.
\newblock {\em \apj} {\bf 2000}, {\em 543},~704--721,
  \href{http://arxiv.org/abs/astro-ph/0006275}{{\normalfont
  [arXiv:astro-ph/astro-ph/0006275]}}.
\newblock {\url{https://doi.org/10.1086/317148}}.

\bibitem[{Athanassoula} and {Misiriotis}(2002)]{AthanassoulaMisiriotis2002}
{Athanassoula}, E.; {Misiriotis}, A.
\newblock {Morphology, photometry and kinematics of N -body bars - I. Three
  models with different halo central concentrations}.
\newblock {\em \mnras} {\bf 2002}, {\em 330},~35--52,
  \href{http://arxiv.org/abs/astro-ph/0111449}{{\normalfont
  [arXiv:astro-ph/astro-ph/0111449]}}.
\newblock {\url{https://doi.org/10.1046/j.1365-8711.2002.05028.x}}.

\bibitem[{Athanassoula}(2003)]{Athanassoula2003}
{Athanassoula}, E.
\newblock {What determines the strength and the slowdown rate of bars?}
\newblock {\em \mnras} {\bf 2003}, {\em 341},~1179--1198,
  \href{http://arxiv.org/abs/astro-ph/0302519}{{\normalfont
  [arXiv:astro-ph/astro-ph/0302519]}}.
\newblock {\url{https://doi.org/10.1046/j.1365-8711.2003.06473.x}}.

\bibitem[{Dubinski} and {Chakrabarty}(2009)]{2009ApJ...703.2068D}
{Dubinski}, J.; {Chakrabarty}, D.
\newblock {Warps and Bars from the External Tidal Torques of Tumbling Dark
  Halos}.
\newblock {\em \apj} {\bf 2009}, {\em 703},~2068--2081,
  \href{http://arxiv.org/abs/0908.0168}{{\normalfont
  [arXiv:astro-ph.GA/0908.0168]}}.
\newblock {\url{https://doi.org/10.1088/0004-637X/703/2/2068}}.

\bibitem[{Villa-Vargas} et~al.(2010){Villa-Vargas}, {Shlosman}, and
  {Heller}]{2010ApJ...719.1470V}
{Villa-Vargas}, J.; {Shlosman}, I.; {Heller}, C.
\newblock {Dark Matter Halos and Evolution of Bars in Disk Galaxies: Varying
  Gas Fraction and Gas Spatial Resolution}.
\newblock {\em \apj} {\bf 2010}, {\em 719},~1470--1480,
  \href{http://arxiv.org/abs/1004.4899}{{\normalfont
  [arXiv:astro-ph.CO/1004.4899]}}.
\newblock {\url{https://doi.org/10.1088/0004-637X/719/2/1470}}.

\bibitem[{Fragkoudi} et~al.(2015){Fragkoudi}, {Athanassoula}, {Bosma}, and
  {Iannuzzi}]{2015MNRAS.450..229F}
{Fragkoudi}, F.; {Athanassoula}, E.; {Bosma}, A.; {Iannuzzi}, F.
\newblock {The effects of Boxy/Peanut bulges on galaxy models}.
\newblock {\em \mnras} {\bf 2015}, {\em 450},~229--245,
  \href{http://arxiv.org/abs/1503.03068}{{\normalfont
  [arXiv:astro-ph.GA/1503.03068]}}.
\newblock {\url{https://doi.org/10.1093/mnras/stv537}}.

\bibitem[{Debattista} et~al.(2017){Debattista}, {Ness}, {Gonzalez}, {Freeman},
  {Zoccali}, and {Minniti}]{Debattista2017}
{Debattista}, V.P.; {Ness}, M.; {Gonzalez}, O.A.; {Freeman}, K.; {Zoccali}, M.;
  {Minniti}, D.
\newblock {Separation of stellar populations by an evolving bar: implications
  for the bulge of the Milky Way}.
\newblock {\em \mnras} {\bf 2017}, {\em 469},~1587--1611,
  \href{http://arxiv.org/abs/1611.09023}{{\normalfont
  [arXiv:astro-ph.GA/1611.09023]}}.
\newblock {\url{https://doi.org/10.1093/mnras/stx947}}.

\bibitem[{Marasco} et~al.(2018){Marasco}, {Oman}, {Navarro}, {Frenk}, and
  {Oosterloo}]{Marasco2018}
{Marasco}, A.; {Oman}, K.A.; {Navarro}, J.F.; {Frenk}, C.S.; {Oosterloo}, T.
\newblock {Bars in dark-matter-dominated dwarf galaxy discs}.
\newblock {\em \mnras} {\bf 2018}, {\em 476},~2168--2176,
  \href{http://arxiv.org/abs/1711.09914}{{\normalfont
  [arXiv:astro-ph.GA/1711.09914]}}.
\newblock {\url{https://doi.org/10.1093/mnras/sty354}}.

\bibitem[{Athanassoula} et~al.(2013){Athanassoula}, {Machado}, and
  {Rodionov}]{Athanassoula2013}
{Athanassoula}, E.; {Machado}, R.E.G.; {Rodionov}, S.A.
\newblock {Bar formation and evolution in disc galaxies with gas and a triaxial
  halo: morphology, bar strength and halo properties}.
\newblock {\em \mnras} {\bf 2013}, {\em 429},~1949--1969,
  \href{http://arxiv.org/abs/1211.6754}{{\normalfont
  [arXiv:astro-ph.CO/1211.6754]}}.
\newblock {\url{https://doi.org/10.1093/mnras/sts452}}.

\bibitem[{{\L}okas} et~al.(2014){{\L}okas}, {Athanassoula}, {Debattista},
  {Valluri}, {Pino}, {Semczuk}, {Gajda}, and {Kowalczyk}]{2014MNRAS.445.1339L}
{{\L}okas}, E.L.; {Athanassoula}, E.; {Debattista}, V.P.; {Valluri}, M.;
  {Pino}, A.d.; {Semczuk}, M.; {Gajda}, G.; {Kowalczyk}, K.
\newblock {Adventures of a tidally induced bar}.
\newblock {\em \mnras} {\bf 2014}, {\em 445},~1339--1350,
  \href{http://arxiv.org/abs/1404.1211}{{\normalfont
  [arXiv:astro-ph.GA/1404.1211]}}.
\newblock {\url{https://doi.org/10.1093/mnras/stu1846}}.

\bibitem[{Lang} et~al.(2014){Lang}, {Holley-Bockelmann}, and
  {Sinha}]{2014ApJ...790L..33L}
{Lang}, M.; {Holley-Bockelmann}, K.; {Sinha}, M.
\newblock {Bar Formation from Galaxy Flybys}.
\newblock {\em \apjl} {\bf 2014}, {\em 790},~L33,
  \href{http://arxiv.org/abs/1405.5832}{{\normalfont
  [arXiv:astro-ph.GA/1405.5832]}}.
\newblock {\url{https://doi.org/10.1088/2041-8205/790/2/L33}}.

\bibitem[{{\L}okas} et~al.(2016){{\L}okas}, {Ebrov{\'a}}, {del Pino},
  {Sybilska}, {Athanassoula}, {Semczuk}, {Gajda}, and {Fouquet}]{Lokas2016}
{{\L}okas}, E.L.; {Ebrov{\'a}}, I.; {del Pino}, A.; {Sybilska}, A.;
  {Athanassoula}, E.; {Semczuk}, M.; {Gajda}, G.; {Fouquet}, S.
\newblock {Tidally Induced Bars of Galaxies in Clusters}.
\newblock {\em \apj} {\bf 2016}, {\em 826},~227,
  \href{http://arxiv.org/abs/1601.07433}{{\normalfont
  [arXiv:astro-ph.GA/1601.07433]}}.
\newblock {\url{https://doi.org/10.3847/0004-637X/826/2/227}}.

\bibitem[{{\L}okas}(2018)]{Lokas2018}
{{\L}okas}, E.L.
\newblock {Formation of Tidally Induced Bars in Galactic Flybys: Prograde
  versus Retrograde Encounters}.
\newblock {\em \apj} {\bf 2018}, {\em 857},~6,
  \href{http://arxiv.org/abs/1803.09465}{{\normalfont
  [arXiv:astro-ph.GA/1803.09465]}}.
\newblock {\url{https://doi.org/10.3847/1538-4357/aab4ff}}.

\bibitem[{Gajda} et~al.(2018){Gajda}, {{\L}okas}, and
  {Athanassoula}]{Gajda2018}
{Gajda}, G.; {{\L}okas}, E.L.; {Athanassoula}, E.
\newblock {Tidally Induced Bars in Gas-rich Dwarf Galaxies Orbiting the Milky
  Way}.
\newblock {\em \apj} {\bf 2018}, {\em 868},~100,
  \href{http://arxiv.org/abs/1807.00674}{{\normalfont
  [arXiv:astro-ph.GA/1807.00674]}}.
\newblock {\url{https://doi.org/10.3847/1538-4357/aaea61}}.

\bibitem[{Scannapieco} et~al.(2009){Scannapieco}, {White}, {Springel}, and
  {Tissera}]{Scannapieco2009}
{Scannapieco}, C.; {White}, S.D.M.; {Springel}, V.; {Tissera}, P.B.
\newblock {The formation and survival of discs in a {\ensuremath{\Lambda}}CDM
  universe}.
\newblock {\em \mnras} {\bf 2009}, {\em 396},~696--708,
  \href{http://arxiv.org/abs/0812.0976}{{\normalfont
  [arXiv:astro-ph/0812.0976]}}.
\newblock {\url{https://doi.org/10.1111/j.1365-2966.2009.14764.x}}.

\bibitem[{Scannapieco} and {Athanassoula}(2012)]{2012MNRAS.425L..10S}
{Scannapieco}, C.; {Athanassoula}, E.
\newblock {Bars in hydrodynamical cosmological simulations}.
\newblock {\em \mnras} {\bf 2012}, {\em 425},~L10--L14,
  \href{http://arxiv.org/abs/1206.5308}{{\normalfont
  [arXiv:astro-ph.CO/1206.5308]}}.
\newblock {\url{https://doi.org/10.1111/j.1745-3933.2012.01291.x}}.

\bibitem[{Algorry} et~al.(2017){Algorry}, {Navarro}, {Abadi}, {Sales}, {Bower},
  {Crain}, {Dalla Vecchia}, {Frenk}, {Schaller}, {Schaye}, and
  {Theuns}]{Algorry_2017}
{Algorry}, D.G.; {Navarro}, J.F.; {Abadi}, M.G.; {Sales}, L.V.; {Bower}, R.G.;
  {Crain}, R.A.; {Dalla Vecchia}, C.; {Frenk}, C.S.; {Schaller}, M.; {Schaye},
  J.;  et~al.
\newblock {Barred galaxies in the EAGLE cosmological hydrodynamical
  simulation}.
\newblock {\em \mnras} {\bf 2017}, {\em 469},~1054--1064,
  \href{http://arxiv.org/abs/1609.05909}{{\normalfont
  [arXiv:astro-ph.GA/1609.05909]}}.
\newblock {\url{https://doi.org/10.1093/mnras/stx1008}}.

\bibitem[{Schaye} et~al.(2015){Schaye}, {Crain}, {Bower}, {Furlong},
  {Schaller}, {Theuns}, {Dalla Vecchia}, {Frenk}, {McCarthy}, {Helly},
  {Jenkins}, {Rosas-Guevara}, {White}, {Baes}, {Booth}, {Camps}, {Navarro},
  {Qu}, {Rahmati}, {Sawala}, {Thomas}, and {Trayford}]{2015MNRAS.446..521S}
{Schaye}, J.; {Crain}, R.A.; {Bower}, R.G.; {Furlong}, M.; {Schaller}, M.;
  {Theuns}, T.; {Dalla Vecchia}, C.; {Frenk}, C.S.; {McCarthy}, I.G.; {Helly},
  J.C.;  et~al.
\newblock {The EAGLE project: simulating the evolution and assembly of galaxies
  and their environments}.
\newblock {\em \mnras} {\bf 2015}, {\em 446},~521--554,
  \href{http://arxiv.org/abs/1407.7040}{{\normalfont
  [arXiv:astro-ph.GA/1407.7040]}}.
\newblock {\url{https://doi.org/10.1093/mnras/stu2058}}.

\bibitem[{Rosas-Guevara} et~al.(2020){Rosas-Guevara}, {Bonoli}, {Dotti},
  {Zana}, {Nelson}, {Pillepich}, {Ho}, {Izquierdo-Villalba}, {Hernquist}, and
  {Pakmor}]{Guevara2020}
{Rosas-Guevara}, Y.; {Bonoli}, S.; {Dotti}, M.; {Zana}, T.; {Nelson}, D.;
  {Pillepich}, A.; {Ho}, L.C.; {Izquierdo-Villalba}, D.; {Hernquist}, L.;
  {Pakmor}, R.
\newblock {The buildup of strongly barred galaxies in the TNG100 simulation}.
\newblock {\em \mnras} {\bf 2020}, {\em 491},~2547--2564,
  \href{http://arxiv.org/abs/1908.00547}{{\normalfont
  [arXiv:astro-ph.GA/1908.00547]}}.
\newblock {\url{https://doi.org/10.1093/mnras/stz3180}}.

\bibitem[{Genel} et~al.(2014){Genel}, {Vogelsberger}, {Springel}, {Sijacki},
  {Nelson}, {Snyder}, {Rodriguez-Gomez}, {Torrey}, and
  {Hernquist}]{2014MNRAS.445..175G}
{Genel}, S.; {Vogelsberger}, M.; {Springel}, V.; {Sijacki}, D.; {Nelson}, D.;
  {Snyder}, G.; {Rodriguez-Gomez}, V.; {Torrey}, P.; {Hernquist}, L.
\newblock {Introducing the Illustris project: the evolution of galaxy
  populations across cosmic time}.
\newblock {\em \mnras} {\bf 2014}, {\em 445},~175--200,
  \href{http://arxiv.org/abs/1405.3749}{{\normalfont
  [arXiv:astro-ph.CO/1405.3749]}}.
\newblock {\url{https://doi.org/10.1093/mnras/stu1654}}.

\bibitem[{Frenk} et~al.(1988){Frenk}, {White}, {Davis}, and
  {Efstathiou}]{Frenk1988}
{Frenk}, C.S.; {White}, S.D.M.; {Davis}, M.; {Efstathiou}, G.
\newblock {The Formation of Dark Halos in a Universe Dominated by Cold Dark
  Matter}.
\newblock {\em \apj} {\bf 1988}, {\em 327},~507.
\newblock {\url{https://doi.org/10.1086/166213}}.

\bibitem[{Dubinski} and {Carlberg}(1991)]{Dubinski1991}
{Dubinski}, J.; {Carlberg}, R.G.
\newblock {The Structure of Cold Dark Matter Halos}.
\newblock {\em \apj} {\bf 1991}, {\em 378},~496.
\newblock {\url{https://doi.org/10.1086/170451}}.

\bibitem[{Warren} et~al.(1992){Warren}, {Quinn}, {Salmon}, and
  {Zurek}]{Warren1992}
{Warren}, M.S.; {Quinn}, P.J.; {Salmon}, J.K.; {Zurek}, W.H.
\newblock {Dark Halos Formed via Dissipationless Collapse. I. Shapes and
  Alignment of Angular Momentum}.
\newblock {\em \apj} {\bf 1992}, {\em 399},~405.
\newblock {\url{https://doi.org/10.1086/171937}}.

\bibitem[{Cole} and {Lacey}(1996)]{Cole1996}
{Cole}, S.; {Lacey}, C.
\newblock {The structure of dark matter haloes in hierarchical clustering
  models}.
\newblock {\em \mnras} {\bf 1996}, {\em 281},~716,
  \href{http://arxiv.org/abs/astro-ph/9510147}{{\normalfont
  [arXiv:astro-ph/astro-ph/9510147]}}.
\newblock {\url{https://doi.org/10.1093/mnras/281.2.716}}.

\bibitem[{Jing}(2002)]{Jing2002}
{Jing}, Y.P.
\newblock {Intrinsic correlation of halo ellipticity and its implications for
  large-scale weak lensing surveys}.
\newblock {\em \mnras} {\bf 2002}, {\em 335},~L89--L93,
  \href{http://arxiv.org/abs/astro-ph/0206098}{{\normalfont
  [arXiv:astro-ph/astro-ph/0206098]}}.
\newblock {\url{https://doi.org/10.1046/j.1365-8711.2002.05899.x}}.

\bibitem[{Bailin} and {Steinmetz}(2005)]{Bailin2005}
{Bailin}, J.; {Steinmetz}, M.
\newblock {Internal and External Alignment of the Shapes and Angular Momenta of
  {\ensuremath{\Lambda}}CDM Halos}.
\newblock {\em \apj} {\bf 2005}, {\em 627},~647--665,
  \href{http://arxiv.org/abs/astro-ph/0408163}{{\normalfont
  [arXiv:astro-ph/astro-ph/0408163]}}.
\newblock {\url{https://doi.org/10.1086/430397}}.

\bibitem[{Allgood} et~al.(2006){Allgood}, {Flores}, {Primack}, {Kravtsov},
  {Wechsler}, {Faltenbacher}, and {Bullock}]{Allgood2006}
{Allgood}, B.; {Flores}, R.A.; {Primack}, J.R.; {Kravtsov}, A.V.; {Wechsler},
  R.H.; {Faltenbacher}, A.; {Bullock}, J.S.
\newblock {The shape of dark matter haloes: dependence on mass, redshift,
  radius and formation}.
\newblock {\em \mnras} {\bf 2006}, {\em 367},~1781--1796,
  \href{http://arxiv.org/abs/astro-ph/0508497}{{\normalfont
  [arXiv:astro-ph/astro-ph/0508497]}}.
\newblock {\url{https://doi.org/10.1111/j.1365-2966.2006.10094.x}}.

\bibitem[{Novak} et~al.(2006){Novak}, {Cox}, {Primack}, {Jonsson}, and
  {Dekel}]{Novak2006}
{Novak}, G.S.; {Cox}, T.J.; {Primack}, J.R.; {Jonsson}, P.; {Dekel}, A.
\newblock {Shapes of Stellar Systems and Dark Halos from Simulations of Galaxy
  Major Mergers}.
\newblock {\em \apjl} {\bf 2006}, {\em 646},~L9--L12,
  \href{http://arxiv.org/abs/astro-ph/0604121}{{\normalfont
  [arXiv:astro-ph/astro-ph/0604121]}}.
\newblock {\url{https://doi.org/10.1086/506605}}.

\bibitem[{Bett} et~al.(2007){Bett}, {Eke}, {Frenk}, {Jenkins}, {Helly}, and
  {Navarro}]{Bett2007}
{Bett}, P.; {Eke}, V.; {Frenk}, C.S.; {Jenkins}, A.; {Helly}, J.; {Navarro}, J.
\newblock {The spin and shape of dark matter haloes in the Millennium
  simulation of a {\ensuremath{\Lambda}} cold dark matter universe}.
\newblock {\em \mnras} {\bf 2007}, {\em 376},~215--232,
  \href{http://arxiv.org/abs/astro-ph/0608607}{{\normalfont
  [arXiv:astro-ph/astro-ph/0608607]}}.
\newblock {\url{https://doi.org/10.1111/j.1365-2966.2007.11432.x}}.

\bibitem[{Velliscig} et~al.(2015){Velliscig}, {Cacciato}, {Schaye}, {Crain},
  {Bower}, {van Daalen}, {Dalla Vecchia}, {Frenk}, {Furlong}, {McCarthy},
  {Schaller}, and {Theuns}]{Velliscig2015}
{Velliscig}, M.; {Cacciato}, M.; {Schaye}, J.; {Crain}, R.A.; {Bower}, R.G.;
  {van Daalen}, M.P.; {Dalla Vecchia}, C.; {Frenk}, C.S.; {Furlong}, M.;
  {McCarthy}, I.G.;  et~al.
\newblock {The alignment and shape of dark matter, stellar, and hot gas
  distributions in the EAGLE and cosmo-OWLS simulations}.
\newblock {\em \mnras} {\bf 2015}, {\em 453},~721--738,
  \href{http://arxiv.org/abs/1504.04025}{{\normalfont
  [arXiv:astro-ph.GA/1504.04025]}}.
\newblock {\url{https://doi.org/10.1093/mnras/stv1690}}.

\bibitem[{Chua} et~al.(2019){Chua}, {Pillepich}, {Vogelsberger}, and
  {Hernquist}]{Chua2019}
{Chua}, K.T.E.; {Pillepich}, A.; {Vogelsberger}, M.; {Hernquist}, L.
\newblock {Shape of dark matter haloes in the Illustris simulation: effects of
  baryons}.
\newblock {\em \mnras} {\bf 2019}, {\em 484},~476--493,
  \href{http://arxiv.org/abs/1809.07255}{{\normalfont
  [arXiv:astro-ph.GA/1809.07255]}}.
\newblock {\url{https://doi.org/10.1093/mnras/sty3531}}.

\bibitem[{Berentzen} and {Shlosman}(2006)]{Berentzen2006}
{Berentzen}, I.; {Shlosman}, I.
\newblock {Growing Live Disks within Cosmologically Assembling Asymmetric
  Halos: Washing Out the Halo Prolateness}.
\newblock {\em \apj} {\bf 2006}, {\em 648},~807--819,
  \href{http://arxiv.org/abs/astro-ph/0603487}{{\normalfont
  [arXiv:astro-ph/astro-ph/0603487]}}.
\newblock {\url{https://doi.org/10.1086/506016}}.

\bibitem[{Abadi} et~al.(2010){Abadi}, {Navarro}, {Fardal}, {Babul}, and
  {Steinmetz}]{Abadi2010}
{Abadi}, M.G.; {Navarro}, J.F.; {Fardal}, M.; {Babul}, A.; {Steinmetz}, M.
\newblock {Galaxy-induced transformation of dark matter haloes}.
\newblock {\em \mnras} {\bf 2010}, {\em 407},~435--446,
  \href{http://arxiv.org/abs/0902.2477}{{\normalfont
  [arXiv:astro-ph.GA/0902.2477]}}.
\newblock {\url{https://doi.org/10.1111/j.1365-2966.2010.16912.x}}.

\bibitem[{Tissera} et~al.(2010){Tissera}, {White}, {Pedrosa}, and
  {Scannapieco}]{Tissera2010}
{Tissera}, P.B.; {White}, S.D.M.; {Pedrosa}, S.; {Scannapieco}, C.
\newblock {Dark matter response to galaxy formation}.
\newblock {\em \mnras} {\bf 2010}, {\em 406},~922--935,
  \href{http://arxiv.org/abs/0911.2316}{{\normalfont
  [arXiv:astro-ph.CO/0911.2316]}}.
\newblock {\url{https://doi.org/10.1111/j.1365-2966.2010.16777.x}}.

\bibitem[{Machado} and {Athanassoula}(2010)]{Machado2010}
{Machado}, R.E.G.; {Athanassoula}, E.
\newblock {Loss of halo triaxiality due to bar formation}.
\newblock {\em \mnras} {\bf 2010}, {\em 406},~2386--2404,
  \href{http://arxiv.org/abs/1004.3874}{{\normalfont
  [arXiv:astro-ph.CO/1004.3874]}}.
\newblock {\url{https://doi.org/10.1111/j.1365-2966.2010.16890.x}}.

\bibitem[{Artale} et~al.(2019){Artale}, {Pedrosa}, {Tissera}, {Cataldi}, and
  {Di Cintio}]{Artale2019}
{Artale}, M.C.; {Pedrosa}, S.E.; {Tissera}, P.B.; {Cataldi}, P.; {Di Cintio},
  A.
\newblock {Dark matter response to galaxy assembly history}.
\newblock {\em \aap} {\bf 2019}, {\em 622},~A197,
  \href{http://arxiv.org/abs/1901.02269}{{\normalfont
  [arXiv:astro-ph.CO/1901.02269]}}.
\newblock {\url{https://doi.org/10.1051/0004-6361/201834096}}.

\bibitem[{Cataldi} et~al.(2021){Cataldi}, {Pedrosa}, {Tissera}, and
  {Artale}]{Cataldi2021}
{Cataldi}, P.; {Pedrosa}, S.E.; {Tissera}, P.B.; {Artale}, M.C.
\newblock {Baryons shaping dark matter haloes}.
\newblock {\em \mnras} {\bf 2021}, {\em 501},~5679--5691,
  \href{http://arxiv.org/abs/2008.02404}{{\normalfont
  [arXiv:astro-ph.GA/2008.02404]}}.
\newblock {\url{https://doi.org/10.1093/mnras/staa3988}}.

\bibitem[{Prada} et~al.(2019){Prada}, {Forero-Romero}, {Grand}, {Pakmor}, and
  {Springel}]{2019MNRAS.tmp.2477P}
{Prada}, J.; {Forero-Romero}, J.E.; {Grand}, R.J.J.; {Pakmor}, R.; {Springel},
  V.
\newblock {Dark matter halo shapes in the Auriga simulations}.
\newblock {\em \mnras} {\bf 2019}, p. 2477,
  \href{http://arxiv.org/abs/1910.04045}{{\normalfont
  [arXiv:astro-ph.GA/1910.04045]}}.
\newblock {\url{https://doi.org/10.1093/mnras/stz2873}}.

\bibitem[{Grand} et~al.(2017){Grand}, {G{\'o}mez}, {Marinacci}, {Pakmor},
  {Springel}, {Campbell}, {Frenk}, {Jenkins}, and {White}]{2017MNRAS.467..179G}
{Grand}, R.J.J.; {G{\'o}mez}, F.A.; {Marinacci}, F.; {Pakmor}, R.; {Springel},
  V.; {Campbell}, D.J.R.; {Frenk}, C.S.; {Jenkins}, A.; {White}, S.D.M.
\newblock {The Auriga Project: the properties and formation mechanisms of disc
  galaxies across cosmic time}.
\newblock {\em \mnras} {\bf 2017}, {\em 467},~179--207,
  \href{http://arxiv.org/abs/1610.01159}{{\normalfont
  [arXiv:astro-ph.GA/1610.01159]}}.
\newblock {\url{https://doi.org/10.1093/mnras/stx071}}.

\bibitem[{Tremaine} and {Weinberg}(1984)]{1984MNRAS.209..729T}
{Tremaine}, S.; {Weinberg}, M.D.
\newblock {Dynamical friction in spherical systems.}
\newblock {\em \mnras} {\bf 1984}, {\em 209},~729--757.
\newblock {\url{https://doi.org/10.1093/mnras/209.4.729}}.

\bibitem[{Weinberg}(1985)]{1985MNRAS.213..451W}
{Weinberg}, M.D.
\newblock {Evolution of barred galaxies by dynamical friction.}
\newblock {\em \mnras} {\bf 1985}, {\em 213},~451--471.
\newblock {\url{https://doi.org/10.1093/mnras/213.3.451}}.

\bibitem[{Hernquist} and {Weinberg}(1992)]{1992ApJ...400...80H}
{Hernquist}, L.; {Weinberg}, M.D.
\newblock {Bar-Spheroid Interaction in Galaxies}.
\newblock {\em \apj} {\bf 1992}, {\em 400},~80.
\newblock {\url{https://doi.org/10.1086/171975}}.

\bibitem[{O'Neill} and {Dubinski}(2003)]{ONeill2003}
{O'Neill}, J.K.; {Dubinski}, J.
\newblock {Detailed comparison of the structures and kinematics of simulated
  and observed barred galaxies}.
\newblock {\em \mnras} {\bf 2003}, {\em 346},~251--264,
  \href{http://arxiv.org/abs/arXiv:astro-ph/0305169}{{\normalfont
  [arXiv:astro-ph/0305169]}}.
\newblock {\url{https://doi.org/10.1046/j.1365-2966.2003.07085.x}}.

\bibitem[{Athanassoula}(2005)]{Athanassoula2005}
{Athanassoula}, E.
\newblock {Dynamical Evolution of Barred Galaxies}.
\newblock {\em Celestial Mechanics and Dynamical Astronomy} {\bf 2005}, {\em
  91},~9--31,  \href{http://arxiv.org/abs/astro-ph/0501196}{{\normalfont
  [arXiv:astro-ph/astro-ph/0501196]}}.
\newblock {\url{https://doi.org/10.1007/s10569-004-4947-7}}.

\bibitem[{Col{\'{\i}}n} et~al.(2006){Col{\'{\i}}n}, {Valenzuela}, and
  {Klypin}]{Colin2006}
{Col{\'{\i}}n}, P.; {Valenzuela}, O.; {Klypin}, A.
\newblock {Bars and Cold Dark Matter Halos}.
\newblock {\em \apj} {\bf 2006}, {\em 644},~687--700,
  \href{http://arxiv.org/abs/astro-ph/0506627}{{\normalfont
  [astro-ph/0506627]}}.
\newblock {\url{https://doi.org/10.1086/503791}}.

\bibitem[{Athanassoula}(2007)]{Athanassoula2007}
{Athanassoula}, E.
\newblock {A bar in the inner halo of barred galaxies - I. Structure and
  kinematics of a representative model}.
\newblock {\em \mnras} {\bf 2007}, {\em 377},~1569--1578,
  \href{http://arxiv.org/abs/astro-ph/0703184}{{\normalfont
  [arXiv:astro-ph/astro-ph/0703184]}}.
\newblock {\url{https://doi.org/10.1111/j.1365-2966.2007.11711.x}}.

\bibitem[{Petersen} et~al.(2016){Petersen}, {Weinberg}, and
  {Katz}]{Petersen2016}
{Petersen}, M.S.; {Weinberg}, M.D.; {Katz}, N.
\newblock {Dark matter trapping by stellar bars: the shadow bar}.
\newblock {\em \mnras} {\bf 2016}, {\em 463},~1952--1967,
  \href{http://arxiv.org/abs/1602.04826}{{\normalfont
  [arXiv:astro-ph.GA/1602.04826]}}.
\newblock {\url{https://doi.org/10.1093/mnras/stw2141}}.

\bibitem[{Holley-Bockelmann} et~al.(2005){Holley-Bockelmann}, {Weinberg}, and
  {Katz}]{HolleyBockelmann2005}
{Holley-Bockelmann}, K.; {Weinberg}, M.; {Katz}, N.
\newblock {Bar-induced evolution of dark matter cusps}.
\newblock {\em \mnras} {\bf 2005}, {\em 363},~991--1007,
  \href{http://arxiv.org/abs/astro-ph/0306374}{{\normalfont
  [astro-ph/0306374]}}.
\newblock {\url{https://doi.org/10.1111/j.1365-2966.2005.09501.x}}.

\bibitem[{Debattista} et~al.(2008){Debattista}, {Moore}, {Quinn},
  {Kazantzidis}, {Maas}, {Mayer}, {Read}, and {Stadel}]{Debattista2008}
{Debattista}, V.P.; {Moore}, B.; {Quinn}, T.; {Kazantzidis}, S.; {Maas}, R.;
  {Mayer}, L.; {Read}, J.; {Stadel}, J.
\newblock {The Causes of Halo Shape Changes Induced by Cooling Baryons: Disks
  versus Substructures}.
\newblock {\em \apj} {\bf 2008}, {\em 681},~1076--1088,
  \href{http://arxiv.org/abs/0707.0737}{{\normalfont [0707.0737]}}.
\newblock {\url{https://doi.org/10.1086/587977}}.

\bibitem[{Athanassoula} et~al.(2005){Athanassoula}, {Ling}, and
  {Nezri}]{2005PhRvD..72h3503A}
{Athanassoula}, E.; {Ling}, F.S.; {Nezri}, E.
\newblock {Halo geometry and dark matter annihilation signal}.
\newblock {\em \prd} {\bf 2005}, {\em 72},~083503,
  \href{http://arxiv.org/abs/astro-ph/0504631}{{\normalfont
  [arXiv:astro-ph/astro-ph/0504631]}}.
\newblock {\url{https://doi.org/10.1103/PhysRevD.72.083503}}.

\bibitem[{Ling} et~al.(2010){Ling}, {Nezri}, {Athanassoula}, and
  {Teyssier}]{2010JCAP...02..012L}
{Ling}, F.S.; {Nezri}, E.; {Athanassoula}, E.; {Teyssier}, R.
\newblock {Dark matter direct detection signals inferred from a cosmological
  N-body simulation with baryons}.
\newblock {\em \jcap} {\bf 2010}, {\em 2010},~012,
  \href{http://arxiv.org/abs/0909.2028}{{\normalfont
  [arXiv:astro-ph.GA/0909.2028]}}.
\newblock {\url{https://doi.org/10.1088/1475-7516/2010/02/012}}.

\bibitem[{Petersen} et~al.(2016){Petersen}, {Katz}, and
  {Weinberg}]{Petersen2016D}
{Petersen}, M.S.; {Katz}, N.; {Weinberg}, M.D.
\newblock {Dynamical response of dark matter to galaxy evolution affects
  direct-detection experiments}.
\newblock {\em \prd} {\bf 2016}, {\em 94},~123013,
  \href{http://arxiv.org/abs/1609.01307}{{\normalfont
  [arXiv:hep-ph/1609.01307]}}.
\newblock {\url{https://doi.org/10.1103/PhysRevD.94.123013}}.

\bibitem[{Kavanagh} and {O'Hare}(2016)]{Kavanagh2016}
{Kavanagh}, B.J.; {O'Hare}, C.A.J.
\newblock {Reconstructing the three-dimensional local dark matter velocity
  distribution}.
\newblock {\em \prd} {\bf 2016}, {\em 94},~123009,
  \href{http://arxiv.org/abs/1609.08630}{{\normalfont [1609.08630]}}.
\newblock {\url{https://doi.org/10.1103/PhysRevD.94.123009}}.

\bibitem[{Machado} and {Manos}(2016)]{Machado_2016}
{Machado}, R.E.G.; {Manos}, T.
\newblock {Chaotic motion and the evolution of morphological components in a
  time-dependent model of a barred galaxy within a dark matter halo}.
\newblock {\em \mnras} {\bf 2016}, {\em 458},~3578--3591,
  \href{http://arxiv.org/abs/1603.02294}{{\normalfont
  [arXiv:astro-ph.GA/1603.02294]}}.
\newblock {\url{https://doi.org/10.1093/mnras/stw572}}.

\bibitem[{Iannuzzi} and {Athanassoula}(2015)]{2015MNRAS.450.2514I}
{Iannuzzi}, F.; {Athanassoula}, E.
\newblock {2D kinematic signatures of boxy/peanut bulges}.
\newblock {\em \mnras} {\bf 2015}, {\em 450},~2514--2538,
  \href{http://arxiv.org/abs/1504.00010}{{\normalfont
  [arXiv:astro-ph.GA/1504.00010]}}.
\newblock {\url{https://doi.org/10.1093/mnras/stv764}}.

\bibitem[{Springel}(2005)]{Springel2005}
{Springel}, V.
\newblock {The cosmological simulation code GADGET-2}.
\newblock {\em \mnras} {\bf 2005}, {\em 364},~1105--1134,
  \href{http://arxiv.org/abs/astro-ph/0505010}{{\normalfont
  [arXiv:astro-ph/astro-ph/0505010]}}.
\newblock {\url{https://doi.org/10.1111/j.1365-2966.2005.09655.x}}.

\bibitem[{Boily} and {Athanassoula}(2006)]{2006MNRAS.369..608B}
{Boily}, C.M.; {Athanassoula}, E.
\newblock {On the equilibrium morphology of systems drawn from spherical
  collapse experiments}.
\newblock {\em \mnras} {\bf 2006}, {\em 369},~608--624,
  \href{http://arxiv.org/abs/0705.2552}{{\normalfont
  [arXiv:astro-ph/0705.2552]}}.
\newblock {\url{https://doi.org/10.1111/j.1365-2966.2006.10365.x}}.

\bibitem[{Pillepich} et~al.(2014){Pillepich}, {Kuhlen}, {Guedes}, and
  {Madau}]{Pillepich2014}
{Pillepich}, A.; {Kuhlen}, M.; {Guedes}, J.; {Madau}, P.
\newblock {The Distribution of Dark Matter in the Milky Way's Disk}.
\newblock {\em \apj} {\bf 2014}, {\em 784},~161,
  \href{http://arxiv.org/abs/1308.1703}{{\normalfont
  [arXiv:astro-ph.GA/1308.1703]}}.
\newblock {\url{https://doi.org/10.1088/0004-637X/784/2/161}}.

\bibitem[{Bozorgnia} et~al.(2016){Bozorgnia}, {Calore}, {Schaller}, {Lovell},
  {Bertone}, {Frenk}, {Crain}, {Navarro}, {Schaye}, and
  {Theuns}]{Bozorgnia2016}
{Bozorgnia}, N.; {Calore}, F.; {Schaller}, M.; {Lovell}, M.; {Bertone}, G.;
  {Frenk}, C.S.; {Crain}, R.A.; {Navarro}, J.F.; {Schaye}, J.; {Theuns}, T.
\newblock {Simulated Milky Way analogues: implications for dark matter direct
  searches}.
\newblock {\em \jcap} {\bf 2016}, {\em 2016},~024,
  \href{http://arxiv.org/abs/1601.04707}{{\normalfont
  [arXiv:astro-ph.CO/1601.04707]}}.
\newblock {\url{https://doi.org/10.1088/1475-7516/2016/05/024}}.

\bibitem[{Kelso} et~al.(2016){Kelso}, {Savage}, {Valluri}, {Freese}, {Stinson},
  and {Bailin}]{Kelso2016}
{Kelso}, C.; {Savage}, C.; {Valluri}, M.; {Freese}, K.; {Stinson}, G.S.;
  {Bailin}, J.
\newblock {The impact of baryons on the direct detection of dark matter}.
\newblock {\em \jcap} {\bf 2016}, {\em 2016},~071,
  \href{http://arxiv.org/abs/1601.04725}{{\normalfont
  [arXiv:astro-ph.GA/1601.04725]}}.
\newblock {\url{https://doi.org/10.1088/1475-7516/2016/08/071}}.

\bibitem[{Sloane} et~al.(2016){Sloane}, {Buckley}, {Brooks}, and
  {Governato}]{Sloane2016}
{Sloane}, J.D.; {Buckley}, M.R.; {Brooks}, A.M.; {Governato}, F.
\newblock {Assessing Astrophysical Uncertainties in Direct Detection with
  Galaxy Simulations}.
\newblock {\em \apj} {\bf 2016}, {\em 831},~93,
  \href{http://arxiv.org/abs/1601.05402}{{\normalfont
  [arXiv:astro-ph.GA/1601.05402]}}.
\newblock {\url{https://doi.org/10.3847/0004-637X/831/1/93}}.

\bibitem[{Lovell} et~al.(2018){Lovell}, {Pillepich}, {Genel}, {Nelson},
  {Springel}, {Pakmor}, {Marinacci}, {Weinberger}, {Torrey}, {Vogelsberger},
  {Alabi}, and {Hernquist}]{2018MNRAS.481.1950L}
{Lovell}, M.R.; {Pillepich}, A.; {Genel}, S.; {Nelson}, D.; {Springel}, V.;
  {Pakmor}, R.; {Marinacci}, F.; {Weinberger}, R.; {Torrey}, P.;
  {Vogelsberger}, M.;  et~al.
\newblock {The fraction of dark matter within galaxies from the IllustrisTNG
  simulations}.
\newblock {\em \mnras} {\bf 2018}, {\em 481},~1950--1975,
  \href{http://arxiv.org/abs/1801.10170}{{\normalfont
  [arXiv:astro-ph.GA/1801.10170]}}.
\newblock {\url{https://doi.org/10.1093/mnras/sty2339}}.

\bibitem[{Evans} et~al.(2019){Evans}, {O'Hare}, and
  {McCabe}]{2019PhRvD..99b3012E}
{Evans}, N.W.; {O'Hare}, C.A.J.; {McCabe}, C.
\newblock {Refinement of the standard halo model for dark matter searches in
  light of the Gaia Sausage}.
\newblock {\em \prd} {\bf 2019}, {\em 99},~023012.
\newblock {\url{https://doi.org/10.1103/PhysRevD.99.023012}}.

\bibitem[{Ibarra} et~al.(2019){Ibarra}, {Kavanagh}, and
  {Rappelt}]{2019JCAP...12..013I}
{Ibarra}, A.; {Kavanagh}, B.J.; {Rappelt}, A.
\newblock {Impact of substructure on local dark matter searches}.
\newblock {\em \jcap} {\bf 2019}, {\em 2019},~013,
  \href{http://arxiv.org/abs/1908.00747}{{\normalfont
  [arXiv:hep-ph/1908.00747]}}.
\newblock {\url{https://doi.org/10.1088/1475-7516/2019/12/013}}.

\bibitem[{Buch} et~al.(2020){Buch}, {Fan}, and {Leung}]{2020PhRvD.101f3026B}
{Buch}, J.; {Fan}, J.; {Leung}, J.S.C.
\newblock {Implications of the Gaia sausage for dark matter nuclear
  interactions}.
\newblock {\em \prd} {\bf 2020}, {\em 101},~063026,
  \href{http://arxiv.org/abs/1910.06356}{{\normalfont
  [arXiv:hep-ph/1910.06356]}}.
\newblock {\url{https://doi.org/10.1103/PhysRevD.101.063026}}.

\bibitem[{Wang} et~al.(2016){Wang}, {Athanassoula}, and
  {Mao}]{2016MNRAS.463.3499W}
{Wang}, Y.; {Athanassoula}, E.; {Mao}, S.
\newblock {Orbital classification in an N-body bar}.
\newblock {\em \mnras} {\bf 2016}, {\em 463},~3499--3512,
  \href{http://arxiv.org/abs/1609.02632}{{\normalfont
  [arXiv:astro-ph.GA/1609.02632]}}.
\newblock {\url{https://doi.org/10.1093/mnras/stw2301}}.

\bibitem[{Manos} and {Machado}(2014)]{Manos_2014}
{Manos}, T.; {Machado}, R.E.G.
\newblock {Chaos and dynamical trends in barred galaxies: bridging the gap
  between N-body simulations and time-dependent analytical models}.
\newblock {\em \mnras} {\bf 2014}, {\em 438},~2201--2217,
  \href{http://arxiv.org/abs/1311.3450}{{\normalfont
  [arXiv:astro-ph.GA/1311.3450]}}.
\newblock {\url{https://doi.org/10.1093/mnras/stt2355}}.

\bibitem[{Miyamoto} and {Nagai}(1975)]{1975PASJ...27..533M}
{Miyamoto}, M.; {Nagai}, R.
\newblock {Three-dimensional models for the distribution of mass in galaxies.}
\newblock {\em \pasj} {\bf 1975}, {\em 27},~533--543.

\bibitem[{Ferrers}(1877)]{Fer1877}
{Ferrers}, N.M.
\newblock {On the Potentials, Ellipsoids, Ellipsoidal Shells, Elliptic Laminae
  and Elliptic Rings, of Variable Densities}.
\newblock {\em \QJPAM} {\bf 1877}, {\em 14},~1--22.

\bibitem[{Pfenniger}(1984)]{Pfenniger1984}
{Pfenniger}, D.
\newblock {The 3D dynamics of barred galaxies}.
\newblock {\em \aap} {\bf 1984}, {\em 134},~373--386.

\end{thebibliography}

\PublishersNote{}
\end{adjustwidth}

\end{document}